\documentclass[reprint,superscriptaddress,amsmath,amssymb,aps,prl]{revtex4-2}
\usepackage{graphicx}
\usepackage{dcolumn}
\usepackage{bm}
\usepackage{color}
\usepackage{units}
\usepackage{wasysym}
\usepackage{MnSymbol}
\usepackage{comment}
\usepackage{times}
\usepackage{soul}

\usepackage[unicode=true,pdfusetitle,bookmarks=false,colorlinks=true,citecolor=blue,urlcolor=blue,linkcolor=blue]{hyperref}
\begin{document}

\title{Theory for the anomalous phase behavior of inertial active Brownian particles}

\author{Jiechao Feng}
\affiliation{Graduate Group in Applied Science \& Technology, University of California, Berkeley, California 94720, USA}
\affiliation{Materials Sciences Division, Lawrence Berkeley National Laboratory, Berkeley, California 94720, USA}
\author{Ahmad K. Omar}
\email{aomar@berkeley.edu}
\affiliation{Department of Materials Science and Engineering, University of California, Berkeley, California 94720, USA}
\affiliation{Materials Sciences Division, Lawrence Berkeley National Laboratory, Berkeley, California 94720, USA}

\begin{abstract}
In contrast to equilibrium systems, inertia can profoundly impact the phase behavior of active systems. 
This has been made particularly evident in recent years, with motility-induced phase separation (MIPS) exhibiting several intriguing dependencies on translational inertia. 
Here we report extensive simulations characterizing the phase behavior of inertial active matter and develop a mechanical theory for the complete phase diagram without appealing to equilibrium notions. 
Our theory qualitatively captures all aspects of liquid-gas coexistence, including the critical value of inertia above which MIPS ceases. 
Notably, our findings highlight that particle softness, and not inertia, is responsible for the MIPS reentrance effect at the center of a proposed active refrigeration cycle.  
\end{abstract}

\maketitle

\textit{Introduction.--} The driving forces for many of the intriguing patterns and phases formed by nonequilibrium systems can entirely differ from those of equilibrium systems. 
Perhaps the most notable example of this is single-component, purely-repulsive active particles which phase separate into two disordered phases despite the absence of attractive interactions~\cite{Fily12,Buttinoni13,Redner13,Cates15}.
This ``motility-induced phase separation'' ~\cite{Fily12,Cates15} (MIPS) has attracted much attention and has served as a model nonequilibrium transition used to generalize thermodynamic arguments.
While the primary axes of the phase diagram that have been interrogated for active particles are density and measures of ``activity'', all nonequilibrium systems inherently have an additional axis as particle inertia can impact the phase behavior~\cite{Lei19,Dai20,Negro22,Shen20,Liao21,Omar23a,Suma14,Mandal2019Motility-InducedPhases,Petrelli18,Petrelli20,Hecht22}. 
This stands in contrast with equilibrium systems where, absent the influence of gravity, particle inertia has no impact on phase behavior and pattern formation. 

While most studies of active systems have focused on the overdamped (i.e.,~inertia-free) limit, Mandal~\textit{et al.} highlighted the significant role of inertia in the phase behavior of two-dimensional (2D) repulsive active disks~\cite{Mandal2019Motility-InducedPhases}.
There, the authors reported that, for finite inertia and for certain fixed density, increasing activity can result in moving from a stable homogeneous system to MIPS and finally \textit{back} to a stable homogeneous system at \textit{large} activities. 
This unexpected reentrance effect, along with the observation that the phases have different kinetic temperatures~\cite{Petrelli20,Mandal2019Motility-InducedPhases}, was used to propose a refrigeration cycle powered by inertial active matter~\cite{Hecht22}.

A number of compelling arguments have been offered to explain some of these intriguing observations~\cite{Suma14,Mandal2019Motility-InducedPhases,Omar23a}; however a theory for the complete inertial dependence of the phase diagram of active matter remains an outstanding challenge.
A mechanical approach to constructing liquid-gas binodals has been proposed that, crucially, makes no appeals to equilibrium ideas~\cite{Aifantis83b,Solon18b, Omar23b}. 
We are thus at an exciting juncture at which many of the anomalous aspects of active phase diagrams may now receive a detailed theoretical treatment. 

In this Letter, we develop a nonequilibrium theory for liquid-gas transition of inertial active systems as a function of density, activity, and inertia.
Using extensive Brownian dynamics simulations of three-dimensional (3D) active spheres, we numerically determine the phase diagram and mechanical equations of state as a function of activity, density, and inertia. 
We analytically derive the nonequilibrium coexistence criteria for active inertial particles \textit{solely in terms of bulk equations} and use our simulations to construct empirical expressions for these state functions.
The resulting theoretical phase diagram qualitatively captures all aspects of the phase diagram, including the increasing (and diverging) critical activity and the nonmonotonicity of the kinetic temperature difference between the phases with increasing inertia. 
We find that the previously observed reentrance effect is not present for active particles in the hard-sphere limit and is in fact driven by the \textit{softness} of the interparticle potential rather than particle inertia. 
More generally, we hope that the framework used in this study can be used to systematically probe the origins and build an intuition for the phase behavior of driven systems as a function of less familiar dynamical control parameters, including inertia. 

\begin{figure*}
	\centering
	\includegraphics[width=.95\textwidth]{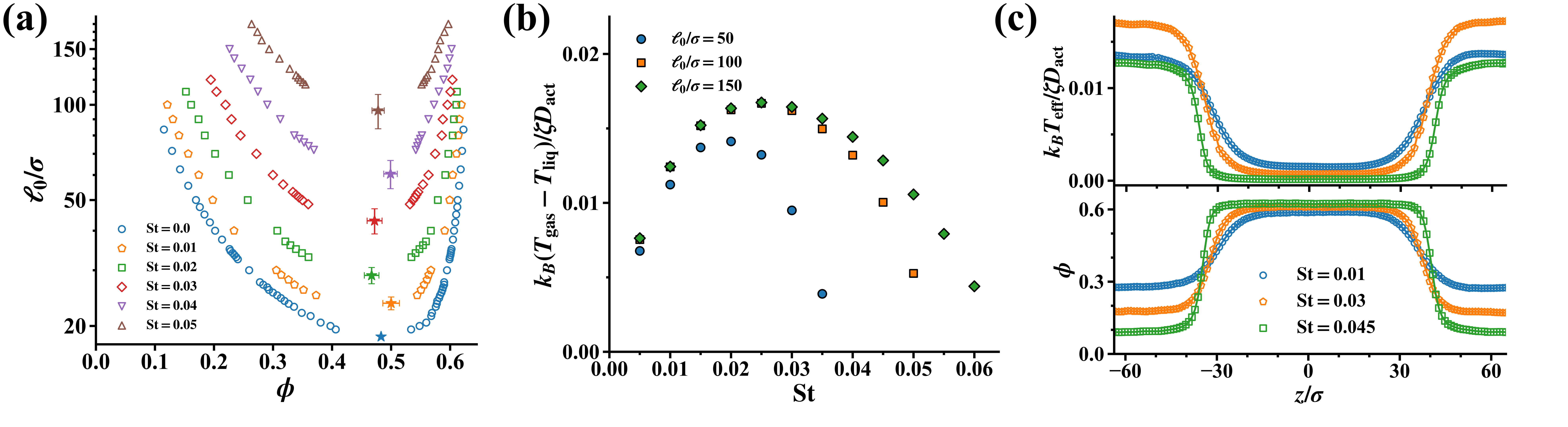}
	\caption{\protect\small{{Simulation results for: (a) binodals of 3D ABPs with ${\rm St}=0.0$ data obtained from Ref.~\cite{Omar21} (data near $\phi=0.5$ are absent due to challenges in obtaining binodal data close to the critical point)}; (b) temperature difference between coexisting phases (where $D_{\rm act}\equiv U_0 \ell_0/6$); and (c) density and kinetic temperature spatial profiles with $\ell_0/\sigma=150.0$ (lines are a guide to the eye).}}
	\label{Fig:1}
\end{figure*}

\textit{Model system.--} We consider the simplest model that captures the inertial effects on MIPS: 3D athermal active spheres~\cite{Omar21} with underdamped translational dynamics. 
Each of the $N$ particles experiences three translational forces: a drag force $-\zeta \dot{\mathbf{x}}$ proportional to the particle velocity $\dot{\mathbf{x}}$ and translational drag coefficient $\zeta$, a conservative interparticle force $\mathbf{F}^{\rm C}[\mathbf{x}^N]$, where $\mathbf{x}^N$ represents all particle positions, and an active force $\mathbf{F}^{\rm A}=\zeta U_0 \mathbf{q}$ where $U_0$ is the intrinsic active speed. 
Stochastic Brownian translational forces, which simply attenuate the influence of activity, are neglected. 
The particle orientation $\mathbf{q}$ independently follow diffusive dynamics $\dot{\mathbf{q}}=\bm{\Omega}\times \mathbf{q}$ where the stochastic angular velocity has mean $\bm{0}$ and variance $\langle \bm{\Omega}(t)\bm{\Omega}(0) \rangle = 2/\tau_{\rm R} \delta(t) \mathbf{{I}}$ and $\tau_{\rm R}$ represents characteristic reorientation time (or inverse rotational diffusion).
Here, we consider only the effects of translational inertia (neglecting rotational inertia) to facilitate a comparison with existing studies~\cite{Mandal2019Motility-InducedPhases,Lowen20,Hecht22,Omar23a}. 

The inclusion of inertia introduces the momentum relaxation time,  $\tau_{\rm M}\equiv m/\zeta$ where $m$ is the particle mass.
The interparticle force $\mathbf{F}^{\rm C}[\mathbf{x}_N; \sigma, \varepsilon]$ is taken from a Weeks-Chandler-Anderson (WCA) potential, which is characterized by a Lennard-Jones diameter $\sigma$ and energy $\varepsilon$~\cite{Weeks71}.
Taking $\zeta U_0$, $\sigma$, and $\tau_{\rm R}$ to be the characteristic units of force, length, and time, respectively, the dimensionless Langevin equation follows as:
\begin{equation}
    \label{eq:eom}
    {\rm St} \ \overline{\ddot{\mathbf{x}}} = - \overline{\dot{\mathbf{x}}} + \frac{\ell_0}{\sigma} \left ( \mathbf{q} + \overline{\mathbf{F}}^{\rm C} [ \overline{\mathbf{x}}^{\rm N}; \mathcal{S}] \right )
\end{equation}
where we have defined the Stokes number ${\rm St}\equiv \tau_{\rm M}/\tau_{\rm R}$ and intrinsic ``run length'' (activity) $\ell_0 \equiv U_0 \tau_{\rm R}$.
The dimensionless force $\overline{\mathbf{F}}^{\rm C}$ depends on the reduced positions $\overline{\mathbf{x}}^{\rm N}$ and is fully characterized by the ``stiffness'' parameter $\mathcal{S} \equiv \varepsilon/(\zeta U_0 \sigma)$~\cite{Omar21}.
For overdamped particles, as $\mathcal{S} \rightarrow \infty$, the finite amplitude of the active force and absence of stochastic translational Brownian forces strictly prevents particles from overlapping within a center-to-center distance of $D = 2^{1/6}\sigma$~\cite{Omar21}.
The particles behave precisely as hard spheres in this limit, despite the use of a continuous potential. 
The system state is fully described by four dimensionless parameters: the volume fraction ($\phi\equiv N\pi D^3/6V$), $\ell_0/\sigma$, ${\rm St}$, and $\mathcal{S}$.

\textit{Simulated phase diagram.--} We use Brownian dynamics simulations to construct the phase diagram with simulation details provided in the Supplemental Material (SM)~\footnote{See Supplemental Material at [URL], which includes Refs.~\cite{Steinhardt83,Speck22,Virtanen20,Paliwal18,Epstein19,Hardy82,Irving50,Lehoucq10,Langford24,Tjhung18,Zakine23}, for additional simulation details, equations of state, the determination of critical points, the discussion about the absence of the MIPS reentrant behavior, the Fokker-Planck analysis to derive the coexistence criteria, and the stability analysis for inertial ABPs.}.
We vary density, activity, and inertia while fixing $\mathcal{S}=50$~\footnote{Our stiffness selection results in an effective hard-sphere diameter of $d_{\rm ex}/D\approx 0.95$, leaving only a narrow range of interparticle separation where continuous repulsions are present.}.
All simulations were conducted with a minimum of $93000$ particles using \texttt{HOOMD-blue}~\cite{Anderson20,Note1}.
The resulting binodal curves are presented in the activity-volume fraction plane for several different values of inertia, shown in Fig.~\ref{Fig:1}(a).
With increasing inertia, the critical activity monotonically increases, while the critical volume fraction displays no statistically significant dependence on inertia.
Here, the critical points are determined by fitting the binodal data nearest to the critical point to power-law critical scaling~\cite{Note1}.
While a comprehensive critical scaling analysis will be required to investigate the critical phenomena of inertial ABPs~\cite{Binder81,Binder87,Rovere88,Rovere90,Rovere93,Siebert18,Partridge19,Maggi21}, the present results make clear that, \textit{for fixed stiffness}, increasing translational inertia monotonically broadens the single-phase region, shifting the binodal to higher activities.

An interesting peculiarity of inertial MIPS is that the kinetic temperature -- defined as $k_{\rm B} T_{\rm eff} = m \langle \dot{\mathbf{x}}^2  \rangle / 3 $ in 3D -- is distinct between the coexisting phases~\cite{Petrelli18,Mandal2019Motility-InducedPhases,Petrelli20,Hecht22}.
The temperature differences between coexisting gas and liquid phases for different ${\rm St}$ and $\ell_0/\sigma$ are displayed in Fig.~\ref{Fig:1}(b) and were extracted from the spatial kinetic temperature profiles (along the axis, $z$, normal to the interface) shown in Fig.~\ref{Fig:1}(c).
For each activity, the inertial dependence of this temperature difference is nonmonotonic and concave.
We can understand the origins of this nonmonotonicity by considering that the kinetic temperature is a decreasing function of density and increasing function of inertia~\cite{Omar23a, Note1}. 
For a fixed activity, increasing inertia initially increases the temperature difference but, eventually, increasing inertia begins to narrow the density difference between the phases as the critical activity shifts to higher values [see Fig.~\ref{Fig:1}(a)] with these competing effects giving rise to the nonmonotonicity. 

\textit{Theory of inertial MIPS.--} 
In the absence of a variational principle, mechanics is a natural choice for constructing nonequilibrium phase diagrams.
The mechanical theory of nonequilibrium fluid-fluid coexistence was recently presented in Ref.~\cite{Omar23b}.
There, a procedure for obtaining the coexistence criteria which makes no appeals to equilibrium notions was outlined.
The criteria that emerge from this mechanical framework can be expressed in the form of a generalized equal-area Maxwell construction. 
Crucially, while the integration variable in this construction is identical for \textit{all equilibrium systems}, out of equilibrium this variable is \textit{system specific} and must be determined from a detailed statistical mechanical treatment. 

\begin{figure*}
	\centering
	\includegraphics[width=.95\textwidth]{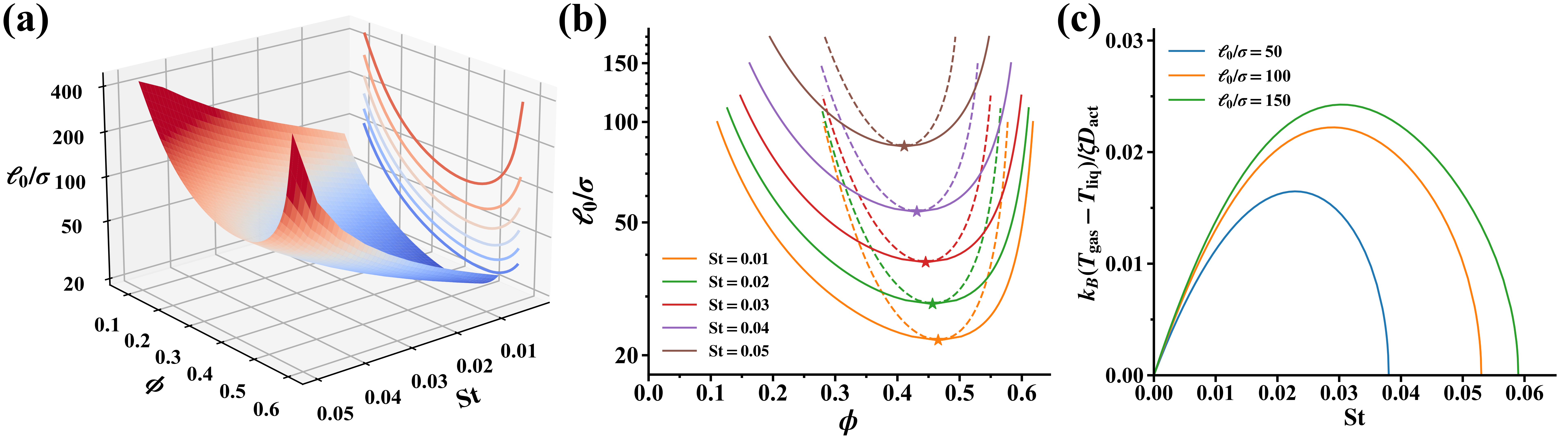}
	\caption{\protect\small{{
	Theoretically predicted: (a) binodals for 3D ABPs in the continuous three-dimensional parameter space; (b) binodals (solid lines) and spinodals (dashed lines) for several ${\rm St}$ (stars denote critical points); and (c) kinetic temperature difference between coexisting phases.}}}
	\label{Fig:2}
\end{figure*}

A statistical mechanical description of interacting ABPs was provided for overdamped active particles in Ref.~\cite{Omar23b}.
Introducing particle inertia results in $3N$ additional degrees of freedom that, unlike equilibrium systems, will have a non-trivial distribution. 
Beginning from our microscopic equation-of-motion [Eq.~\eqref{eq:eom}], the probability density $f_N(\bm{\Gamma}; t)$ of finding the system in a microstate $\bm{\Gamma}=(\mathbf{x}^N, \dot{\mathbf{x}}^N, \mathbf{q}^N)$ satisfies $\partial f_N / \partial t = \mathcal{L} f_N$, where $\mathcal{L}$ is a dynamical operator specific to the equation-of-motion~\cite{Note1}.
In the case of phase separation the order parameter is the number density field $\rho(\mathbf{x}; t)= \langle \hat{\rho}(\mathbf{x}) \rangle  = \langle \sum_{\alpha =1 }^N \delta(\mathbf{x}-\mathbf{x}_{\alpha}) \rangle$ with its evolution equation given by $\partial \rho / \partial t = \int_{\gamma} \hat{\rho} \mathcal{L} f_N d \bm{\Gamma}$, where $\langle ... \rangle = \int (...) f_N (\bm{\Gamma},t) d \bm{\Gamma} $ is the ensemble average and $\gamma$ denotes the phase-space volume.

We now simplify our formally exact expressions using the conditions of stationary phase coexistence with a planar interface and taking $z$ to be the direction normal to the interface. 
Solving for the density profile of a quasi one-dimensional flux-free steady state requires solving the static mechanical balance, $d\Sigma_{zz}/dz=0$~\cite{Fily17,Omar23b}, where $\bm{\Sigma}$ is the \textit{dynamic stress} tensor which includes both the effects of true stresses and body forces~\cite{Omar20, Omar23a}.
Straightforward integration of the mechanical balance under these conditions reveals that the dynamic stress must be spatially uniform, with $\Sigma_{zz} = C$.
When the dynamic stress can be expressed as:
\begin{equation}
\label{eq:Mech_Theory_const_stress_equil}
     - \Sigma_{zz} = \mathcal{P}(\rho) - a(\rho) \frac{d^2\rho}{dz^2} - b(\rho) \left(\frac{d\rho}{dz}\right)^2 \ ,
\end{equation}
where $\mathcal{P}$ is the bulk dynamic pressure and $a$ and $b$ represent interfacial coefficients, the condition of uniform stress results in the following nonequilibrium coexistence criteria~\cite{Omar23a}: 
\begin{subequations}
\begin{align}
    \mathcal{P}(\rho_{\rm liq}) = \mathcal{P}(\rho_{\rm gas}) &= \mathcal{P}^{\rm coexist} \ , \label{eq:eq_pressure} \\
    \int_{\rho_{\rm gas}}^{\rho_{\rm liq}} \left [ \mathcal{P}(\rho) - \mathcal{P}^{\rm coexist}\right  ] E(\rho) d\rho &= 0  \label{eq:int_pressure} \ ,
\end{align}
where
\begin{equation}
\label{eq:E_general}
    E (\rho) = \frac{1}{a(\rho)}   \exp\left( 2 \int \frac{b(\rho)} {a(\rho)} \ d\rho \right) \ .
\end{equation}
\end{subequations}
The second criterion~\eqref{eq:int_pressure}, is a \textit{weighted-area} Maxwell construction in the $\mathcal{P}-\rho$ plane where $E(\rho)$ is a weighting factor [Eq.~\eqref{eq:E_general}] defined such that $\rho(z)$ does not appear in the criterion: only bulk equations of state ($\mathcal{P}$, $a$, and $b$) are needed to now find the binodal densities~\footnote{This weighted-area construction is identical to an equal-area Maxwell construction in the $\mathcal{P}-\mathcal{E}$ plane where $\partial \mathcal{E}/\partial\rho = E$.}.

To use this nonequilibrium coexistence framework we must: (i) formally expand our exact expression of the dynamic stress to match the form of Eq.~\eqref{eq:Mech_Theory_const_stress_equil}; and (ii) construct equations of state for the terms appearing in the expansion~\cite{Solon15}. 
The dynamic stress for interacting inertial active particles is found to have three distinct contributions arising from particle interactions, the active force, and the kinetic energy of the particles~\cite{Note1}. 
Discarding higher-order gradient terms and gradient terms with coefficients that are independent or weakly dependent on activity, we are able to cast our dynamic stress in the desired form, finding (in 3D)~\cite{Note1}:
\begin{subequations}
\begin{align}
    \mathcal{P}(\rho) =& \  p_{\rm int}+p_{\rm act}+p_{\rm  k}  \ , \label{eq:P} \\
    a(\rho) =&   \frac{\ell_0^2 }{20 (1+6 {\rm St})(1+2 {\rm St})} \Big[ \frac{6 {\rm St}}{ 1+ {\rm St}} +1 \Big] \overline{U}^2 \frac{\partial p_{\rm int}  }{\partial  \rho }  \label{eq:a}  \ , \\
    b(\rho) =&   \frac{\ell_0^2 }{20 (1+6 {\rm St})(1+2 {\rm St})}  \Big[  \frac{6 {\rm St}}{ 1+ {\rm St}} \frac{\partial }{\partial \rho } \Big( \overline{U}^2 \frac{\partial p_{\rm int}  }{\partial \rho} \Big)  \nonumber \\
    &+ \overline{U} \frac{\partial}{\partial \rho} \Big( \overline{U} \frac{\partial p_{\rm int}  }{\partial \rho}\Big)  \Big]   \label{eq:b}  \ ,
\end{align} 
where $\overline{U} \equiv (p_{\rm act}+p_{\rm k}) /(\zeta D^{\rm act}\rho)$ is a dimensionless active speed with $\overline{U} \in [0,1]$. 
We thus identify the weighting factor:
\begin{equation}
    E(\rho) = \overline{U}^{12 {\rm St}/(1+7{\rm St})} \frac{\partial p_{\rm int} }{\partial \rho} \label{eq:E} \ ,
\end{equation}
\end{subequations}
and can now formally construct the liquid-gas binodal for inertial active particles \textit{without appealing to equilibrium notions}. 
We directly fit our equations of state ($p_{\rm int}$, $p_{\rm act}$, $p_{\rm k}$) to homogeneous fluid simulation data while ensuring that all known  limits are enforced.
The resulting analytical forms of $p_{\rm k}$, $p_{\rm act}$, and $p_{\rm int}$ with respect to $(\phi, \ell_0/\sigma, {\rm St})$ are provided in the SM~\cite{Note1}.
Moreover, as detailed in the SM~\cite{Note1}, several closures and approximations were made to derive Eqs.~\eqref{eq:P}-\eqref{eq:E}. 
While these assumptions may affect the quantitative accuracy of the coexistence criteria, our goal is to qualitatively understand how inertia alters these criteria compared to overdamped ABPs. 
We believe that the key finding—that inertia has a highly nonlinear effect on the coexistence criteria—is likely robust.

Employing our coexistence criteria and equations of state, we construct the three-dimensional liquid-gas binodal surface for inertial ABPs across the entire parameter space $(\ell_0/\sigma, \phi, {\rm St})$ [see Fig.~\ref{Fig:2}(a)].
Slices of the binodal for several values of inertia are presented in Fig.~\ref{Fig:2}(b) alongside the spinodal, the latter of which was obtained from a mechanical linear stability analysis~\cite{Note1}.
The theoretical binodal correctly captures the trend of an increasing  critical activity and, intriguingly, predicts a systematic reduction in the critical density with increasing inertia.
As presented in Fig.~S9~\cite{Note1}, the binodal of inertial ABPs can be well approximated by the overdamped theory \textit{despite} the nonlinear dependence of the coexistence criteria on inertia. 
This nonlinear dependence is not sufficient to quantitatively alter binodal from the overdamped prediction precisely because MIPS occurs at relatively small St and the coexistence theory for inertial ABPs reduces to that of Ref.~\cite{Omar23b} as ${\rm St}\to 0$.
We can also predict the kinetic temperature difference between coexisting phases as shown in Fig.~\ref{Fig:2}(c), correctly capturing the qualitative trend of a nonmonotonic inertia dependence found in our simulations. 

\begin{figure}
	\centering
	\includegraphics[width=.475\textwidth]{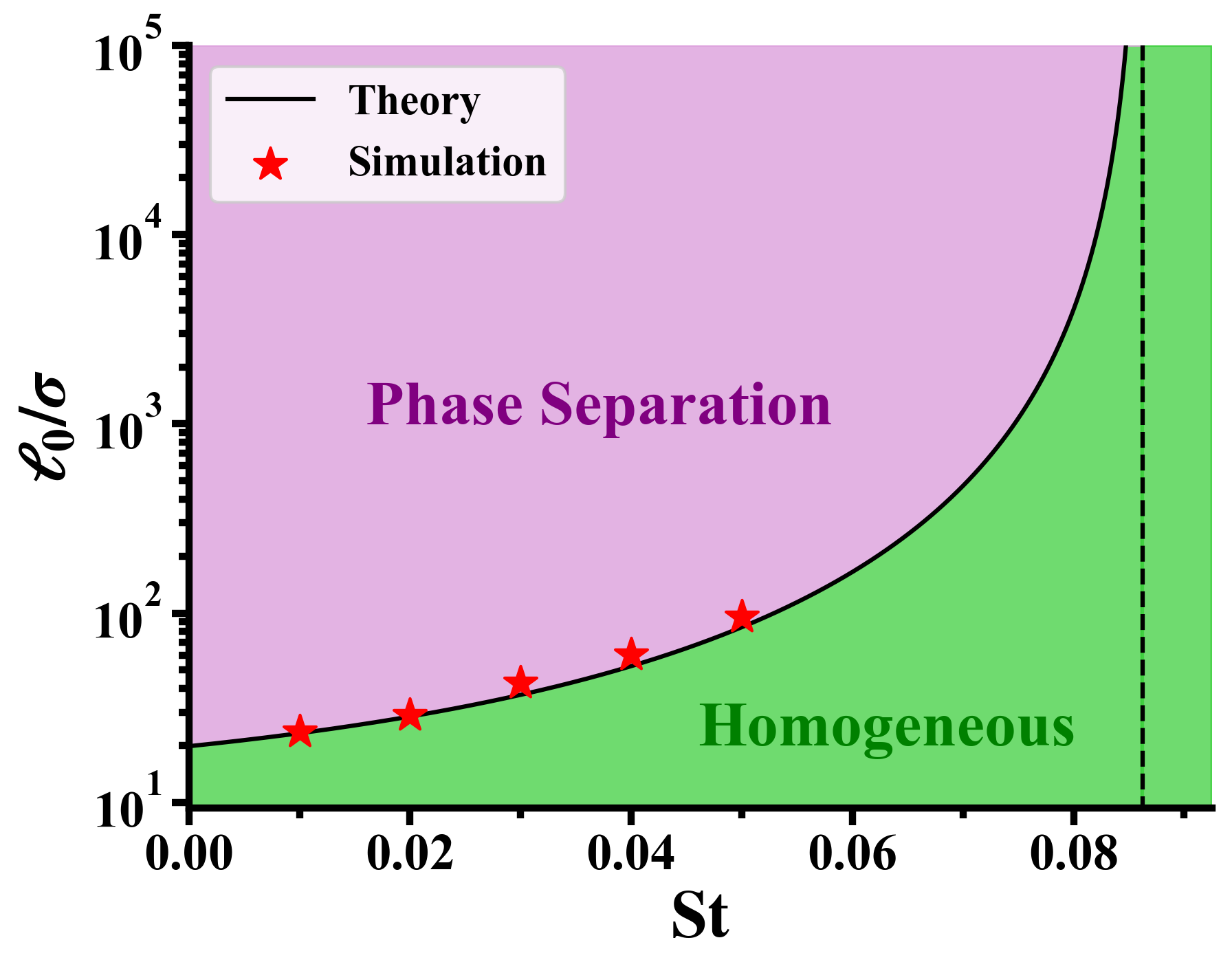}
	\caption{\protect\small{{
	Theoretically predicted critical activity with ${\rm St}$. Stars denote simulation results. The vertical dashed line is located at ${\rm St}=0.086$.}}}
	\label{Fig:3}
\end{figure}

Our theory predicts that the critical activity diverges as ${\rm St} \to 0.086$ (see Fig.~\ref{Fig:3}).
For ${\rm St}$ larger than this incredibly small value of inertia, MIPS is entirely suppressed for all activities and densities.
While the precise value of this critical ${\rm St}$ depends on the specific forms of equations of state, its existence is consistent with our simulations, which also aligns with earlier findings~\cite{Suma14, Mandal2019Motility-InducedPhases, Omar23a}.
Moreover, Fig.~\ref{Fig:3} makes clear that there is no value of inertia where increasing activity can result in a phase-separated system to homogenize, the reentrance effect reported in Ref.~\cite{Mandal2019Motility-InducedPhases} is absent.

We can understand the inertia-induced stability of active fluids by considering the stability criteria for phase separation. 
A homogeneous fluid with a dynamic pressure that decreases with increasing density (resulting in a van der Waals loop), $\partial \mathcal{P}/\partial \rho<0$, is unstable with respect to linear density perturbations~\cite{Note1}. 
In the absence of inertia, the dynamic pressure of active fluids with purely repulsive interactions consists of $p_{\rm int}$ and $p_{\rm act}$, the latter of which can exhibit a nonmonontonic density dependence and induce MIPS at high activities.
While inertia also generates $p_{\rm k}$, straightforward energy conservation arguments result in the relation $p_{\rm k}/p_{\rm act}=2  {\rm St}$ in 3D~\cite{Omar23a}: \textit{both} the kinetic and active pressures can destabilize active fluids.

\begin{figure*}
	\centering
	\includegraphics[width=.95\textwidth]{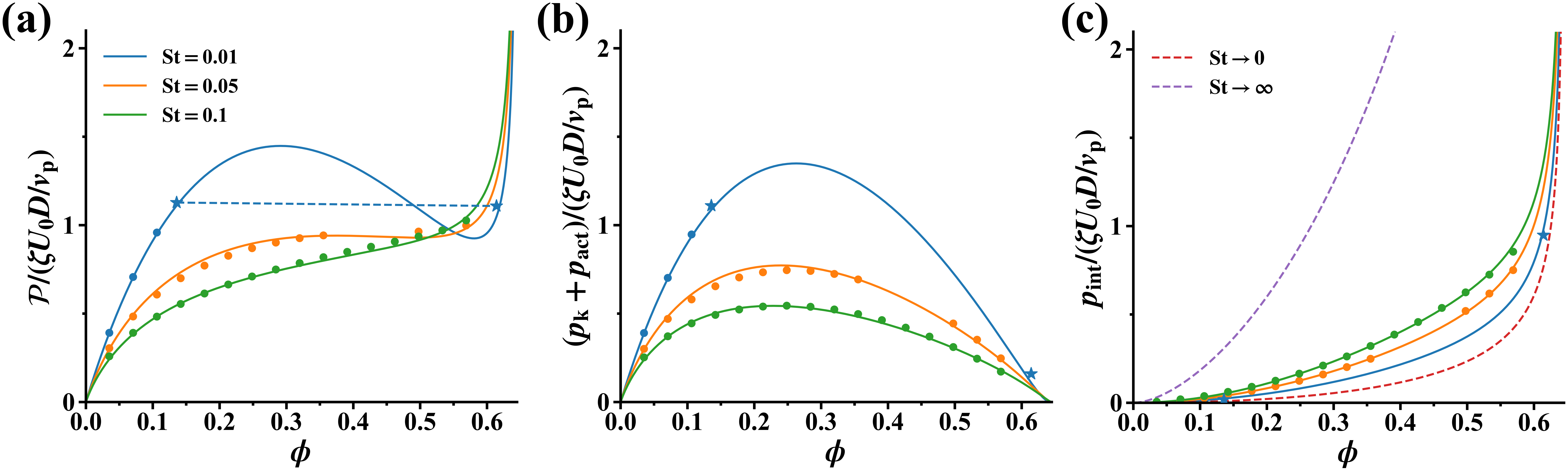}
	\caption{\protect\small{{
	Inertia dependence of: (a) $\mathcal{P}$; (b)  $p_{\rm k}+p_{\rm act}$; and (c) $p_{\rm int}$ for $\ell_0/\sigma=85.0$. Solid lines indicate fits while symbols denote simulation results. In (a), the dashed line and stars respectively denote $\mathcal{P}^{\rm coexist}$ and the binodal densities from simulations with ${\rm St}=0.01$. In (c), the red and purple dashed lines respectively denote the predicted asymptotic forms of $p_{\rm int}$ as ${\rm St}\to 0$ and ${\rm St}\to \infty$. $v_{\rm p}\equiv\pi D^3/6$ is the particle volume.}}}
	\label{Fig:4}
\end{figure*}

Figure~\ref{Fig:4}(a) displays the density dependence of our dynamic pressure as a function of inertia for a fixed activity. 
Points on the curves correspond to data obtained from simulations of homogeneous fluids (the absence of data is indicative of regions where MIPS quickly occurs).
At the lowest value of inertia, a prominent van der Waals loop exists in $\mathcal{P}$ signifying a fluid that is unstable for a broad range of densities~\footnote{Examining the individual components of the pressure, we see as expected that $p_{\rm k} + p_{\rm act}$ is destabilizing for a broad range of densities while the interaction pressure, $p_{\rm int}$, monotonically increases with density and is thus stabilizing for all densities. In fact, at the ``random close packing'' density of $\phi_{\rm RCP} \approx 0.645$,  $p_{\rm k} + p_{\rm act}$ must vanish due to the immobility of active particles while $p_{\rm int}$ diverges.}.
With increasing inertia, the van der Waals loop rapidly diminishes and is eventually eliminated entirely. 
We can see that the increased stability arises from the reduced magnitude of the contributions of  $p_{\rm k} + p_{\rm act}$ relative to those of $p_{\rm int}$. 
The increasing interaction pressure with inertia reflects the enhanced probability of interparticle contacts arising from the sluggish particle dynamics: particles collide and remain in close proximity for a longer duration. 
These inertia-enhanced collisions reduce the effective speed of the particles and thus the active pressure.
In contrast, the kinetic pressure continues to increase with inertia for all densities. 
This increase is not enough to compensate for the reduced active pressure and increased interaction pressure, resulting in a closing of the van der Waals loop with increasing inertia and stable homogeneous active fluids.  

\textit{Origins of reentrant MIPS.--}
Mandal~\textit{et al.}~\cite{Mandal2019Motility-InducedPhases} first observed a reentrance effect, in which a crossover from a homogeneous system to MIPS and back to a homogeneous system was observed with increasing activity for a certain range of inertia.
The origins of this observation (and its connection to particle inertia) have remained, to the best of our knowledge, unclear.
We now identify the exact origin of this reentrance effect.
Often, varying the activity of ABPs is done by increasing the active force while keeping all other parameters fixed (including the interaction parameters).
As a result, while the dimensionless activity $\ell_0/\sigma$ increases with increasing active force the particle stiffness $\mathcal{S}$ was simultaneously inversely decreasing: the particles become effectively softer. 
This increasing softness allows particles to more closely approach each other and achieve higher densities. 
The origins of this reentrance effect is thus not due to inertia, but rather, a consequence of the simultaneous variation of both activity and interparticle stiffness. 
In other words, the basic physics rooted in the reentrance of MIPS is the increasingly unhindered particle motion due to the diminished role of interparticle repulsion with decreasing stiffness. 
The slowing down of particles due to particle interactions is a crucial ingredient in MIPS and the reduction of this effect with increasing softness is the origin of MIPS reentrance.

Fixing $\mathcal{S}$ while varying $\ell_0/\sigma$ is precisely why the reentrance effect was not observed in our work -- its origins are rooted in $\mathcal{S}$ rather than ${\rm St}$. 
We have explicitly verified that a variable $\mathcal{S}$ is required for a reentrant MIPS using both 2D and 3D inertial ABPs.
Notably, in \textit{overdamped} 2D ABPs, the reentrance effect persists when $\mathcal{S}$ decreases with increasing $\ell_0/\sigma$, reinforcing that a variable $\mathcal{S}$ drives this effect~\cite{Note1, Bialke13}.
Indeed, previous work has reported the reentrance of MIPS in overdamped self-propelled repulsive disks~\cite{Bialke13} and others have speculated that the observed reentrant effect is an effect of variable particle stiffness~\cite{Stenhammer14}.
Reentrance appears to be found in all repulsive systems exhibiting MIPS if $\mathcal{S}$ is varied but is entirely absent when $\mathcal{S}$ is held constant.

\textit{Conclusions.--} 
In this work, we have developed a theory for the full phase diagram of active Brownian spheres as a function of activity, translational inertia, and density. 
Our theory is able to reproduce all qualitative trends observed in our simulations while also offering an intuitive  picture for why inertia eliminates MIPS (particles collide more and move less). 
Moreover, we find the reentrance effect that forms the basis of a recently proposed active refrigeration cycle, has origins that are rooted in particle interaction softness. 
As inertia can play a prominent role in the dynamics of both living and artificial active systems, it is our hope that the theory presented in this work could be useful to understand various experimental phenomena exhibited by these diverse systems~\cite{Patterson17,Scholz18a,Scholz18b,Deblais18,Mayya19,Sprenger23,Fersula24}.
For example, from the experimental parameters in Ref.~\cite{Scholz18a}, we determined a Stokes number ${\rm St} = 0.0184$ for phase-separating rotating robots, which is just below the critical value where MIPS is expected to break down and thus consistent with our theory.
More generally, we hope that the framework used in this work can be extended to interrogate additional axes of the active phase diagram, including rotational inertia~\cite{Farhadi18,Kokot17,Su21, Lisin22,Bayati22,Zhang22,Caprini22}, and include forms of coexistence with order parameters beyond the density.  

\begin{acknowledgments}
This work was supported by the Laboratory Directed Research and Development Program of Lawrence Berkeley National Laboratory under U.S. Department of Energy Contract No. DE-AC02-05CH11231.
J.F. acknowledges support from the UC Berkeley College of Engineering Jane Lewis Fellowship. 
This research used the Savio computational cluster resource provided by the Berkeley Research Computing program. 
The data that support the findings of this study are available from the corresponding author upon reasonable request.
\end{acknowledgments}

\end{document}


\title{Supplemental Material -- Theory for the anomalous phase behavior of inertial active Brownian particles}

\author{Jiechao Feng}
\affiliation{Graduate Group in Applied Science \& Technology, University of California, Berkeley, California 94720, USA}
\affiliation{Materials Sciences Division, Lawrence Berkeley National Laboratory, Berkeley, California 94720, USA}
\author{Ahmad K. Omar}
\email{aomar@berkeley.edu}
\affiliation{Department of Materials Science and Engineering, University of California, Berkeley, California 94720, USA}
\affiliation{Materials Sciences Division, Lawrence Berkeley National Laboratory, Berkeley, California 94720, USA}

\maketitle
\vspace{-5mm}
\tableofcontents

\newpage
\setcounter{page}{0}
\section{Additional Simulation Details}
\subsection{Simulation Settings in 3D}
All 3D simulations were conducted for a minimum duration of $2\times 10^4 \sigma/U_0$ with $93311$ particles.
For these simulations, we extract the coexisting liquid and gas densities by conducting ``slab'' simulations – constant volume simulations with one box dimension larger than the other two ($L_z>L_x=L_y$).
This box geometry yields a one-dimensional density profile along the $z$-axis, accompanied by a well-defined interface between the coexisting liquid and gas phases.
We use a box aspect ratio of $L_z/L_x \approx 4$ in most of our simulations.
Our total volume fraction $\phi$ is adjusted with $\ell_0/\sigma$ to be approximately the average of the anticipated liquid and gas phase densities.
This ensures that the system is within the spinodal and MIPS is spontaneous while also leading to a significant amount (by volume) of both phases. 
We thus extract the coexisting liquid and gas densities by fitting the one-dimensional density profile to a sigmoidal function (see Fig.~1(c) in the main text for an illustration)~\cite{Omar21}:
\begin{equation}
    \phi(z) = \frac{\phi_{\rm liq}-\phi_{\rm gas}}{2} \tanh \left (\frac{z-z_0}{w} \right) + \frac{\phi_{\rm liq}+\phi_{\rm gas}}{2} \ ,
    \label{eq:fit}
\end{equation}
where $\phi_{\rm liq}$ and $\phi_{\rm gas}$ are the liquid and gas phase densities, $w$ is the interfacial width, and $z_0$ represents a horizontal shift.

As we approach the critical point, obtaining a density profile that can be approximately fit using Eq.~\eqref{eq:fit} becomes more challenging due to the diminishing difference between coexisting densities.
Therefore, for these simulations, we initially prepare the system at a volume fraction $\phi \gtrsim \phi_{\rm liq}$ and subsequently perform a uniaxial elongation to achieve our target density.
Such a procedure enables us to obtain an improved density profile with a single liquid and gas domain, which can then be fit using Eq.~\eqref{eq:fit}~\cite{Omar20,Omar21}.

\subsection{Simulation Settings in 2D}
We used a minimum of $40000$ particles for all 2D simulations.
For 2D simulations, we observed that it is difficult to stabilize a single liquid (or gas) ``slab'' and are thus unable to fit the density profile to Eq.~\eqref{eq:fit}.
Instead, we extract the coexisting densities by computing the local Voronoi volume (area) of each particle, $V^{\rm Vor}_i$.
The local area fraction of an individual particle is $\phi_i = \pi  D^2/(4 V^{\rm Vor}_i)$.
We identify the liquid and gas phase densities by locating the peaks in the bimodal distribution of $\phi$. 
\newpage

\subsection{Interaction Potential}
\begin{figure}[h]
	\centering
	\includegraphics[width=.4\textwidth]{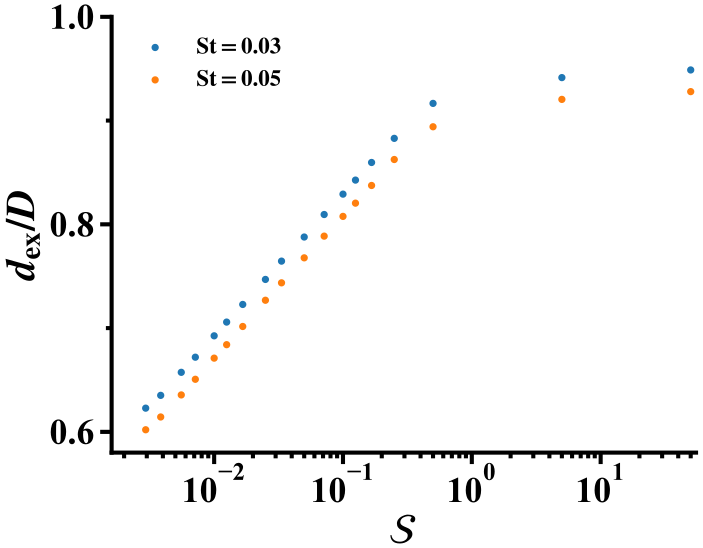}
	\caption{\protect\small{{Minimum pair separation distance $d_{\rm ex}$ of different stiffness ${\mathcal S}$ for ${\rm St}=0.03$ and ${\rm St}=0.05$.}}}
	\label{Fig:S1}
\end{figure}
As stated in the main text, the ``stiffness'' parameter $\mathcal{S} \equiv \varepsilon/(\zeta U_0 \sigma)$ determines whether hard-sphere statistics is recovered for active spheres interacting with a WCA potential~\cite{Weeks71}.
To understand this, note that the force between a pair of particles can be expressed as $\mathbf{F}^{\rm C}_{ij}(\mathbf{x}_{ij})=-\bm{\nabla} \varepsilon u^{\rm WCA}(r;\sigma)$ where $\mathbf{x}_{ij}$ is the distance between a pair of particles and the dimensionless potential $u^{\rm WCA}$ has the following form~\cite{Weeks71}:
\begin{equation*}
u^{\rm WCA}(r;\sigma)= 
\begin{cases}
  4\left[ (\frac{\sigma}{r})^{12}-(\frac{\sigma}{r})^6 \right] +1, \ & r\le D \\
  0, & r > D
\end{cases}
\end{equation*}
where $D=2^{1/6} \sigma$. 
Selecting $\zeta U_0$ and $\sigma$ to be the units of force and length, respectively, results in a dimensionless force $\overline{\mathbf{F}}^{\rm C}_{ij}(\overline{\mathbf{x}}_{ij}; \mathcal{S})=\mathcal{S} \overline{\bm{\nabla}} u^{\rm WCA} (\overline{r})$ (where $\overline{\bm{\nabla}} \equiv \bm{\nabla}\sigma$ is the dimensionless gradient operator) that is entirely characterized by the stiffness parameter $\mathcal{S}$.
A selection of $\mathcal{S}=50$ ensures that the active force cannot generate overlaps within a pair separation distance, $d_{\rm ex}$, of $d_{\rm ex}/D \approx 0.9997$, leaving a negligible range where continuous repulsions are present.
While the finite amplitude of the active force results in $d_{\rm ex} \approx D$ acting identically to a hard-sphere diameter for $\mathcal{S}=50$ \textit{for overdamped dynamics}, finite inertia can result in particle overlaps that reduce the effective diameter, $d_{\rm ex}$.
By computing the radial distribution function of homogeneous active fluids, $g(r)$, we can determine the separation distance in which particles do not overlap.
Fig.~\ref{Fig:S1} displays the dependence of $d_{\rm ex}$ (as determined through the solution to the equation $g(r=d_{\rm ex})=0.01$) as a function of $\mathcal{S}$ and ${\rm St}$ for particles in 2D.
While inertia is found to ``soften'' particles with increasing inertia, we can also clearly see that we are approaching the hard-sphere limit with increasing stiffness. 
A chosen timestep of $5\times 10^{-4} \sigma/U_0$ ensures minimal particle overlap and results in effective hard-sphere statistics in our simulation.
Practically, we set  $\mathcal{S} = 50$ which results in $d_{\rm ex}/D \approx 0.95$. 
While there is a narrow range of particle separations in which continuous repulsions are present, hard-sphere statistics remain closely approximated. 

\section{Equations of State for Inertial ABPs}
We directly fit the equations of state for the three contributions to the dynamic pressure ($p_k$, $p_{\rm act}$, $p_{\rm int}$) from simulations within regions of the $(\phi, \ell_0/\sigma, {\rm St})$ parameter space where the system remains a homogeneous fluid.
The microscopic definition~\cite{Omar23a} of the pressure contributions are:
\begin{align*}
    p_{\rm k} &= \rho m \langle \dot{\mathbf{x}} \cdot \dot{\mathbf{x}} \rangle /d \ ,\\
    p_{\rm act} &= \rho \tau_{\rm R} \langle \dot{\mathbf{x}} \cdot \mathbf{F}^{\rm A} \rangle / d(d-1) \ , \\
    p_{\rm int} &= \rho \langle \mathbf{x} \cdot \mathbf{F}^{\rm C} \rangle /d \ ,
\end{align*}
where $\langle \cdot \rangle$ denotes an average over all particles and $d$ denotes the spatial dimension.
We conducted simulations for different $\ell_0/\sigma$ and ${\rm St}$ for $0<\phi<\phi_{\rm RCP}$ ($\phi_{\rm RCP}=0.645$ is the volume fraction of random close packing).
When $\phi$ approaches $\phi_{\rm RCP}$, the system may crystallize~\cite{Omar21, Omar23a}.
As we aim to focus on the liquid-gas phase boundary, we require equations of state solely for homogeneous fluids. 
We therefore exclude data points from crystalline systems by measuring the system-averaged per-particle Steinhardt-Nelson-Ronchetti order parameter $\langle q_{12} \rangle$~\cite{Steinhardt83}
\begin{equation*}
    q_{l, i} = \left( \frac{4\pi}{2l+1}\sum \limits_{m=-l}^l |\langle Y_{lm, i} \rangle |^2 \right)^{1/2} \ ,
\end{equation*}
where $\langle Y_{lm, i} \rangle$ is the average spherical harmonics of the bond angles formed between particle $i$ and its nearest neighbors.
We take $\langle q_{12} \rangle >0.3$ to indicate the presence of an active solid and discard this data from our fits.

Our goal is to fit $p_{\rm k}$, $p_{\rm act}$, and $p_{\rm int}$ as functions of $\phi$, $\ell_0/\sigma$, and ${\rm St}$.
Elementary energy conservation arguments discussed in Ref.~\cite{Omar23a} provide the relation $p_{\rm k}/p_{\rm act}=(d-1){\rm St}$.
Our aim is thus reduced to fitting two independent equations of state: $p_{\rm k}+p_{\rm act}$ and $p_{\rm int}$. 
For $p_{\rm k}+p_{\rm act}$, inspired by the functional form of $p_{\rm act}$ for overdamped active spheres proposed in Ref.~\cite{Omar23b}, we propose the following form of $p_{\rm k}+p_{\rm act}$:
\begin{equation}
    \label{eq:pk_pact}
    \frac{p_{\rm k}+p_{\rm act}}{\zeta U_0 D /v_{\rm p}} = \frac{\phi}{6} \left (\frac{\ell_0}{D} \right ) \exp{  \left[ - \frac{A \phi^B}{(1-\phi/\phi_{\rm RCP})^C } \right] } ,
\end{equation}
where $A$, $B$, and $C$ are functions of $\ell_0/\sigma$ and ${\rm St}$ that are to be fit and $v_{\rm p}=\pi D^3/6$ is the volume of a single particle.

For $p_{\rm int}$ we propose:
\begin{equation}
    \label{eq:pint}
    \frac{p_{\rm int}}{\zeta U_0 D /v_{\rm p}} = \frac{E\phi^2 + F \phi^3 + G\phi^4}{(1-\phi/\phi_{\rm RCP})^{0.486}} \ ,
\end{equation}
where $E$, $F$, and $G$ are functions of $\ell_0/\sigma$ and ${\rm St}$ and the exponent 0.486 is obtained from fitting. 
Note that we have enforced that $p_{\rm int}$ diverges as $\phi$ approaches $\phi_{\rm RCP}$.
From the simulation results, the empirically fit functions are
\begin{align*}
    A(\ell_0/\sigma, {\rm St}) &= \left[ \tanh (0.287 \ln (\ell_0/\sigma) -1.611 ) +1 \right]  \left[ 90.793 \sinh (0.156{\rm St}) \exp (-2.930 {\rm St}) + 3.576 \right] ,  \\
    B(\ell_0/\sigma, {\rm St}) &= 1.131 \tanh (-6.185 \ln (\ell_0/\sigma) - 10.416 ) \sinh (16.005 {\rm St}) \exp (-15.688 {\rm St} ) + 1.049 , \\
    C(\ell_0/\sigma, {\rm St}) &= 0.847\tanh (0.186 \ln (\ell_0/\sigma) - 0.369 ) \frac{ \exp (-29.482 {\rm St}) - \exp (28.473 {\rm St} ) }{ \exp (-29.482 {\rm St}) + \exp (28.473 {\rm St}) } + 0.554 , \\
    E (\ell_0/\sigma, {\rm St}) &= \left[ 0.077  (\ln (\ell_0/\sigma))^2 - 0.046 \ln (\ell_0/\sigma) + 0.098 \right] \left[ 0.187 (\ln {\rm St} )^2 +1.805 \ln {\rm St} + 5.275 \right] , \\
    F (\ell_0/\sigma, {\rm St}) &= \left[ 0.377  (\ln (\ell_0/\sigma))^2 - 0.230 \ln (\ell_0/\sigma) - 1.333 \right] \left[ -0.083 (\ln {\rm St} )^2 -0.782 \ln {\rm St} -2.079 \right] , \\
    G (\ell_0/\sigma, {\rm St}) &= \left[ -0.123  (\ln (\ell_0/\sigma))^2 + 0.180 \ln (\ell_0/\sigma) + 1.410 \right] \left[ -0.131 (\ln {\rm St} )^2 - 1.261 \ln {\rm St} -2.757 \right] .
\end{align*}

Here we elaborate on the process of deriving these functional forms.
Firstly, Eqs. (S2) and (S3) are designed to capture the correct physical limits and trends of the corresponding equations of state.
For instance, when $\ell_0/\sigma \to 0$, $p_{\rm k}+p_{\rm act}$ should approach the reversible limit $p_{\rm k}+p_{\rm act} = \rho k_B T_{\rm act}$ for all $\phi < \phi_{\rm RCP}$ as the equation-of-motion for the particles will satisfy the fluctuation-dissipation theorem with $k_B T_{\rm act}=\zeta U_0 \ell_0/6$ serving as an effective temperature~\cite{Omar21, Omar23a}. 
This leads us to to enforce $A\to 0$ in this reversible limit~\cite{Evans23}.
Moreover, when ${\rm St}$ increases from $0$ to $\infty$, for fixed $\ell_0/\sigma$ and ${\rm St}$, $p_{\rm int}$ should monotonically increase from the overdamped limit to the equilibrium hard-sphere limit~\cite{Omar23b,Evans23}:
\begin{equation*}
    \frac{p_{\rm int}}{\zeta U_0 D /v_{\rm p}} ({\rm St \to 0}) = 2^{-7/6} \frac{\phi^2}{\sqrt{1-\phi/\phi_{\rm RCP}}} \ , \\
\end{equation*}
\begin{equation*}
    \frac{p_{\rm int}}{\zeta U_0 D /v_{\rm p}} ({\rm St \to \infty}) = \frac{4 k_B T_{\rm eff}}{\zeta U_0 D} \frac{\phi^2}{(1-\phi/\phi_{\rm RCP})^{0.76}}  \sum \limits_{n=0}^8 c_n 4^n \phi^n \ , 
\end{equation*}
where $k_B T_{\rm eff}$ is the corresponding effective temperature at ${\rm St \to \infty}$.

Next, the functional forms of $A$, $B$, ... , $G$, are designed by considering specific physical limits and trends.
For example, we anticipate that the effects of $\ell_0/\sigma$ and ${\rm St}$ to be largely independent, so most of these functional forms are the product of one function of $\ell_0/\sigma$  and another function of ${\rm St}$.
We used Python package \texttt{scipy.optimize.curve\_fit()} to perform a multi-dimensional fit with independent variables $\phi$, $\ell_0/\sigma$, and ${\rm St}$ to determine the numerical coefficients in $A$, $B$, ... , $G$~\cite{Virtanen20}.
If the initial fit results were unsatisfactory, we adjusted these functional forms, iterating until the error between our fits and data was reduced below some tolerance.

\begin{figure}[h]
	\centering
	\includegraphics[width=.95\textwidth]{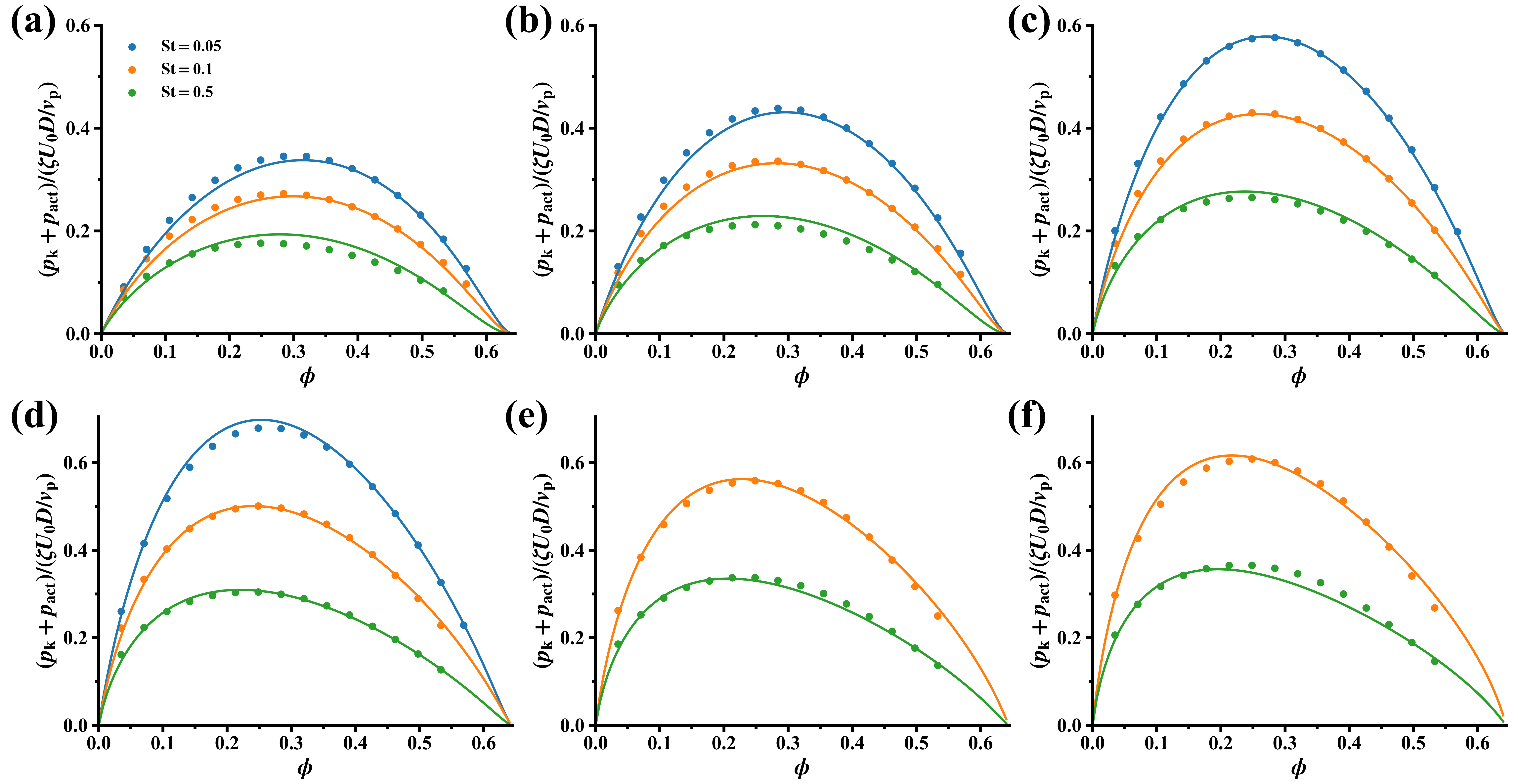}
	\caption{\protect\small{{Fits of $p_{\rm k}+p_{\rm act}$ for different $\ell_0/\sigma$ and ${\rm St}$, denoted by solid lines. Symbols represent simulation results. The corresponding activities $\ell_0/\sigma$ are (a) $20.0$, (b) $30.0$, (c) $50.0$, (d) $70.0$, (e) $90.0$, and (f) $110.0$, respectively.}}}
	\label{Fig:S2}
\end{figure}

\begin{figure}[h]
	\centering
	\includegraphics[width=.95\textwidth]{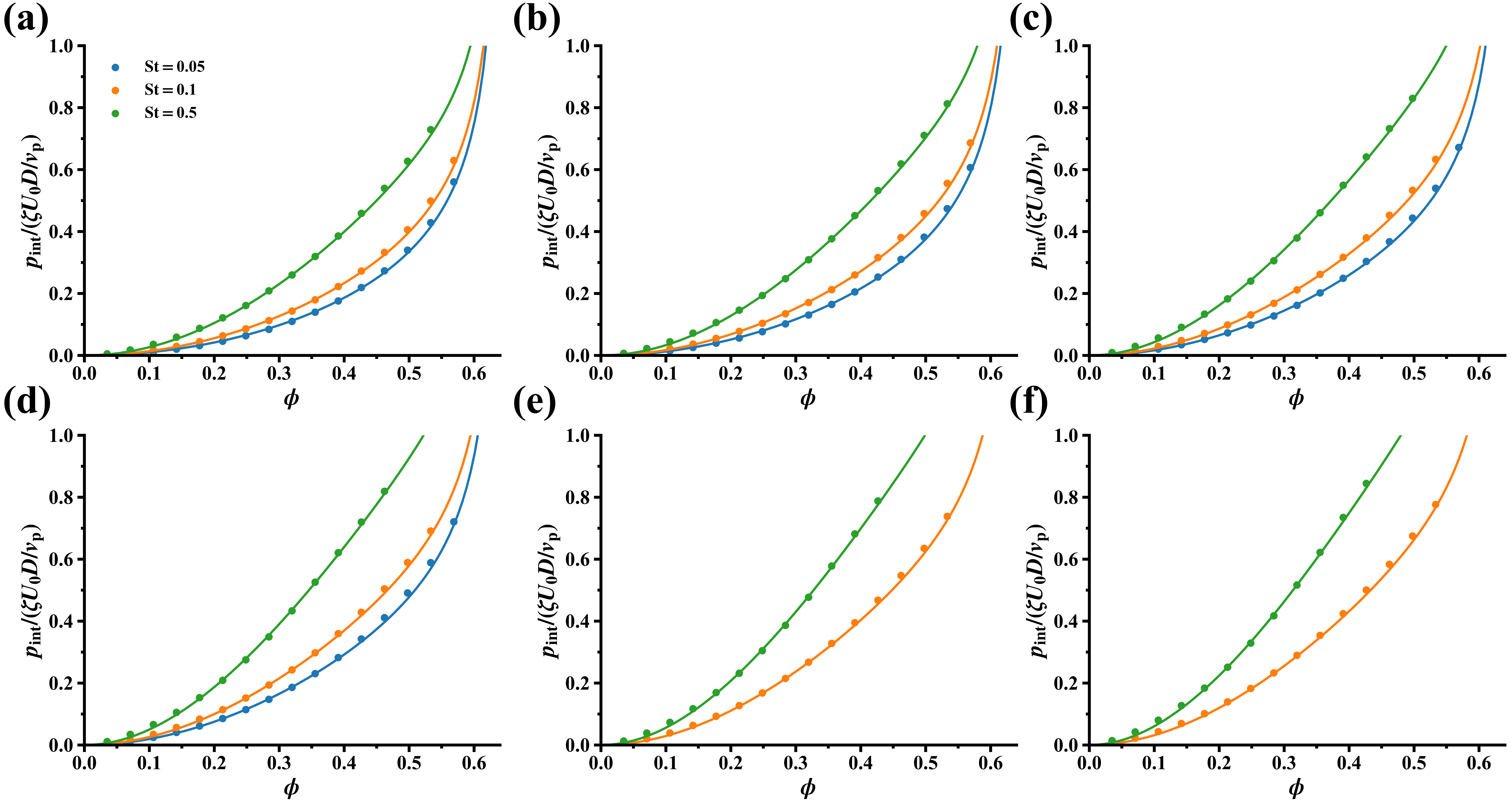}
	\caption{\protect\small{{Fits of $p_{\rm int}$ for different $\ell_0/\sigma$ and ${\rm St}$, denoted by solid lines. Symbols represent simulation results. The corresponding activities $\ell_0/\sigma$ are (a) $20.0$, (b) $30.0$, (c) $50.0$, (d) $70.0$, (e) $90.0$, and (f) $110.0$, respectively.}}}
	\label{Fig:S3}
\end{figure}

Figures~\ref{Fig:S2} and ~\ref{Fig:S3} display our equations of state alongside the simulation data.
Although we needed to fit approximately 20 coefficients in the functional forms of $p_{\rm k}
+p_{\rm act}$ and $p_{\rm int}$, we avoided overfitting by using a much larger data set, with around 250 data points for the fitting process.
For Fig.~4(c) in the main text, we approximate the effective temperature in the infinite inertia limit using ${\rm St} = 10$ as the asymptotic form of  $k_B T_{\rm eff}$, which appears to have been reached at these finite values of inertia. 
The coefficients $c_n$ are provided in Table~\ref{table1}.
\begin{table}[h]
\caption{Coefficients in $p_{\rm int}({\rm St \to \infty})$ obtained from Ref.~\cite{Song89}}
\begin{ruledtabular}
\scriptsize
\begin{tabular}{cccccccccc}
$n$    & $0$ & $1$                   & $2$                   & $3$                   & $4$        &  $5$                    & $6$                    & $7$                   & $8$                \\ 
$c_n$  & $1$ & $1.649\times 10^{-1}$ & $2.217\times 10^{-2}$ & $1.840\times 10^{-3}$ & $3.373\times 10^{-5}$  &  $-1.117\times 10^{-5}$ & $-8.914\times 10^{-6}$ & $-9.469\times 10^{-7}$ & $-4.356\times 10^{-7}$ 
\end{tabular}
\end{ruledtabular}
\label{table1}
\end{table}

\pagebreak

\section{Determination of Critical Points}
\begin{figure}[h]
	\centering
	\includegraphics[width=.94\textwidth]{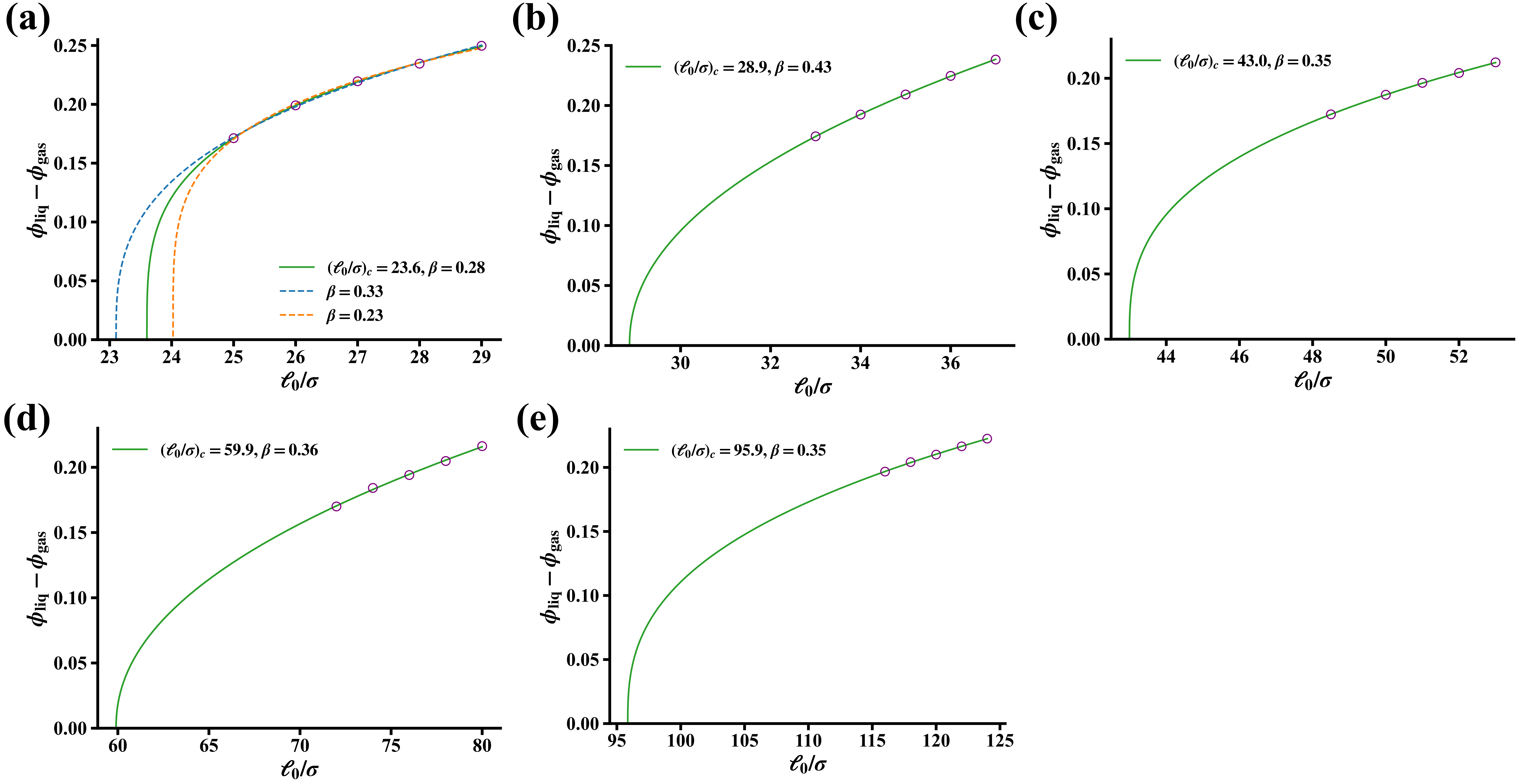}
	\caption{\protect\small{{
	Fits of the critical exponent $\beta$ for ${\rm St}=$ (a) $0.01$, (b) $0.02$, (c) $0.03$, (d) $0.04$, and (e) $0.05$, respectively. Circles denote simulation data, which were the only ones used in the corresponding critical scaling analysis.}}}
	\label{Fig:S4}
\end{figure}

\begin{figure}[h]
 	\centering
	\includegraphics[width=.94\textwidth]{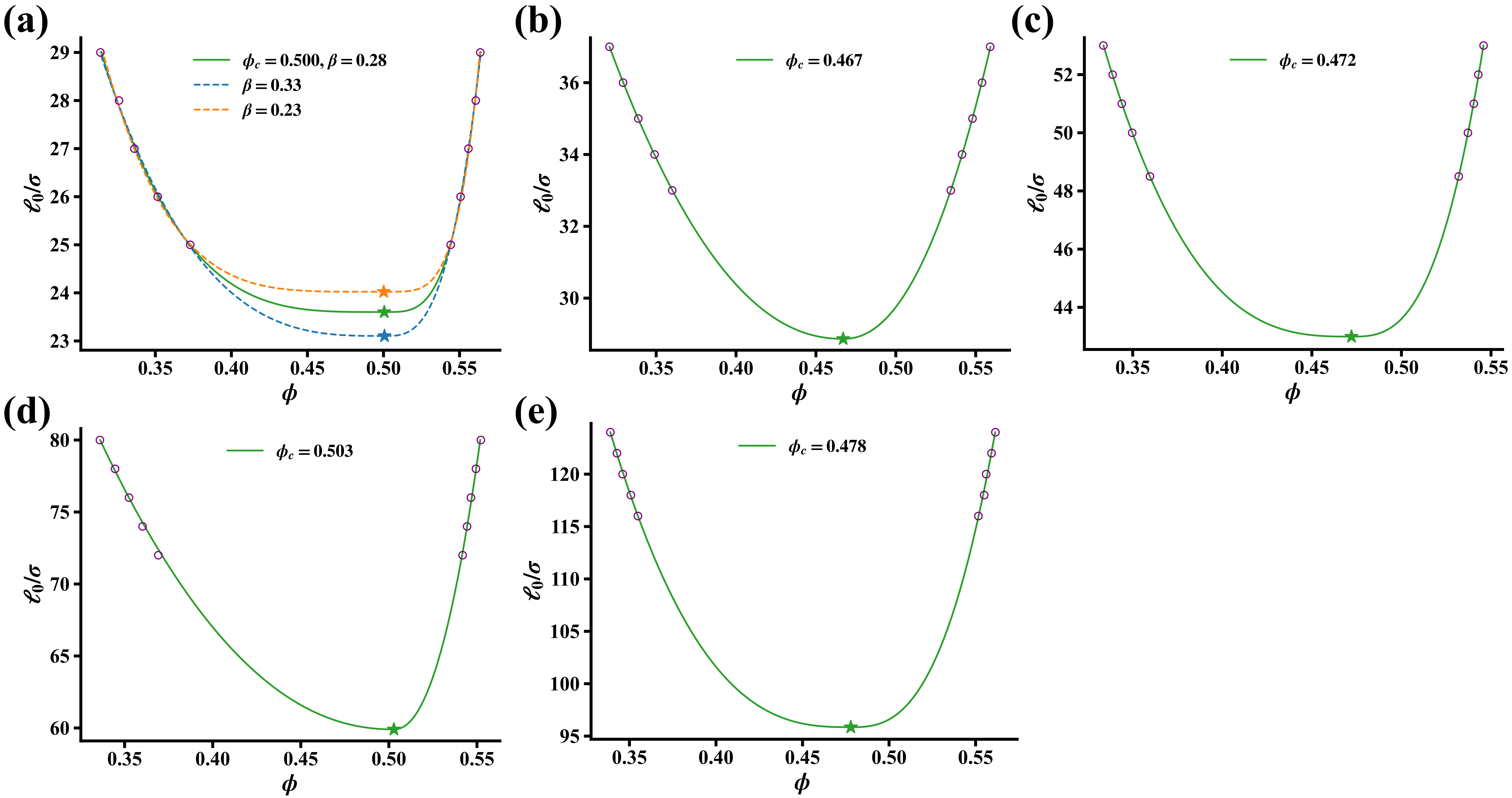}
	\caption{\protect\small{{
	Fits of the critical volume fraction $\phi_{\rm c}$ for ${\rm St}=$ (a) $0.01$, (b) $0.02$, (c) $0.03$, (d) $0.04$, and (e) $0.05$, respectively. Circles denote simulation data, while stars denote the critical point.}}}
	\label{Fig:S5}
\end{figure}

The critical point of MIPS could be rigorously determined through established approaches. 
For example, the critical activity can be identified as the activity where the Binder parameter exhibits scale invariance~\cite{Binder81,Binder87,Rovere88,Rovere90,Rovere93}.
This method enables an independent determination of the critical activity as well as the resulting critical exponent.
Notably, this technique has recently been applied to active systems~\cite{Siebert18,Partridge19,Maggi21}.
While we hope to conduct a full Binder analysis in the future, we use a critical scaling ansatz to estimate the location of the critical point in our simulation~\cite{Omar21}.
Specifically, we define the reduced intrinsic run length as
\begin{equation*}
    \tau \equiv \frac{l_0/\sigma-(l_0/\sigma)_{\rm c}}{(l_0/\sigma)_{\rm c}} \ ,
\end{equation*}
and the order parameter as
\begin{equation*}
    \Delta \phi \equiv \phi_{\rm liq}-\phi_{\rm gas} \ .
\end{equation*}
We expect that 
\begin{equation*}
    \tau = A \Delta \phi ^{\beta} \ ,
\end{equation*}
where $A$ is only a function of ${\rm St}$. 
As this scaling is only anticipated near the critical point ($\tau \ll 1$), we perform this analysis on our simulation data nearest to the critical point.
We simultaneously fit $(\ell_0/\sigma)_{\rm c}$ and $\beta$ to this data and subsequently use the resulting fits to extract $\phi_c$.
The outcomes of this two-step process are presented in Fig.~\ref{Fig:S4} and Fig.~\ref{Fig:S5}. 

Our study reveals that the critical exponents $\beta$ fall within the range of $0.28 \le \beta \le 0.46$ with an average of approximately $0.36$.
This aligns closely with the finding of Ref.~\cite{Omar21}, which reported $\beta \approx 0.33$, consistent with the 3D Ising universality class.
The categorization of the critical behavior of 2D ABPs within 2D Ising universality class is currently a topic of debate~\cite{Siebert18,Partridge19,Maggi21,Speck22}.
Our multiparameter fit, while not precise enough to definitively determine the value of $\beta$ [Fig.~\ref{Fig:S4}(a) and Fig.~\ref{Fig:S5}(a)], suggests that an independent determination of critical activity (such as through Binder analysis), could significantly reduce the uncertainty surrounding the critical exponent.
This would further our understanding of the critical behavior of 3D ABPs with inertia.
The error bars associated with the critical points shown in Fig.~1(a) in the main text were directly obtained from the covariance matrix obtained during our fitting procedure~\cite{Virtanen20}.

\pagebreak

\section{Absence of the MIPS Reentrant Behavior}
\begin{figure}[h]
	\centering
	\includegraphics[width=.95\textwidth]{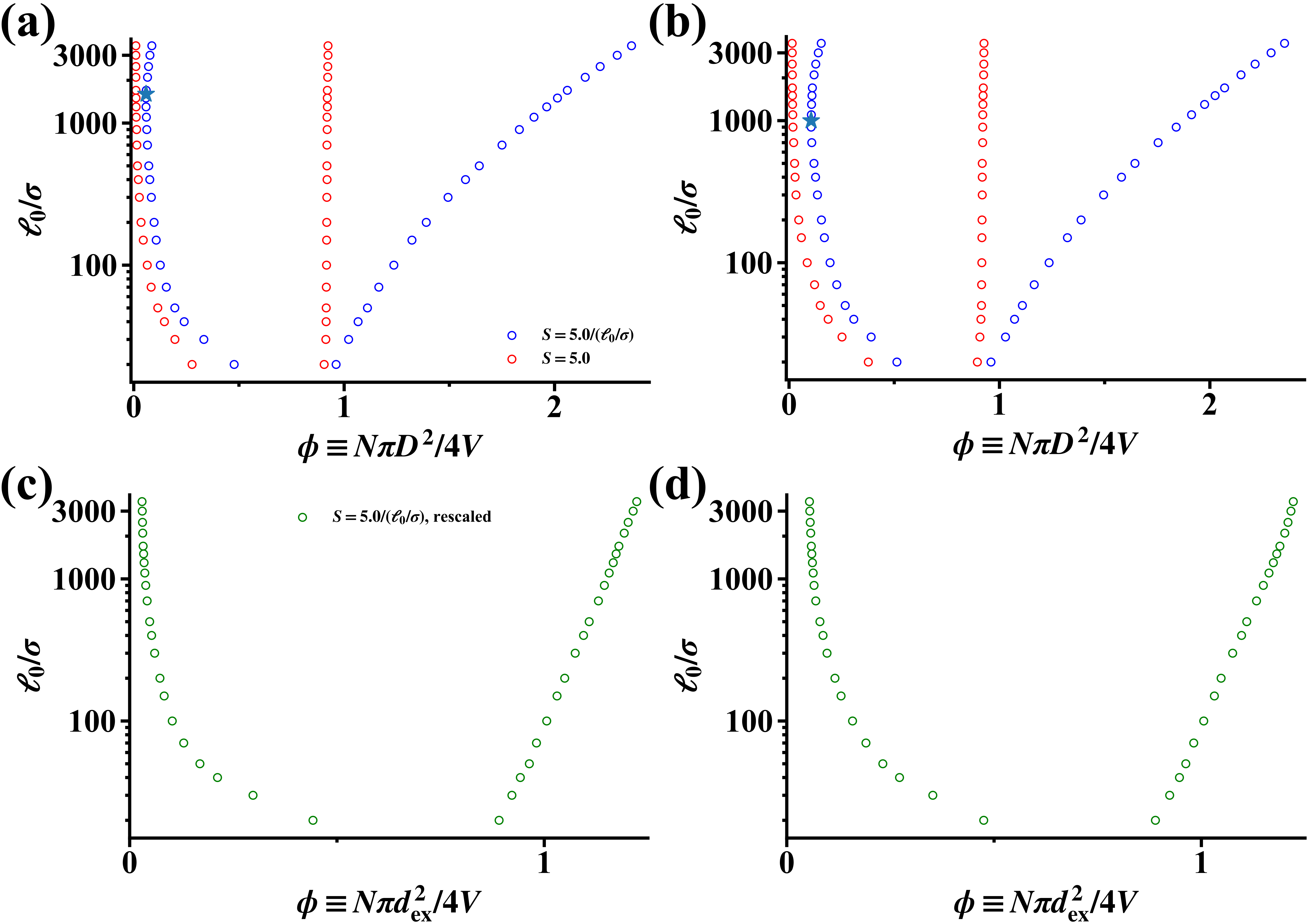}
	\caption{\protect\small{{Simulated binodal for 2D inertial ABPs. In (a) ${\rm St}=0.03$ and (b) $0.05$, blue circles denote varying stiffness with $\mathcal{S}=5.0/(\ell_0/\sigma)$, red circles denote fixed stiffness $\mathcal{S}=5.0$.  Blue stars denote the critical activity where the gas phase density begins to increase with $\ell_0/\sigma$. In (c) ${\rm St}=0.03$ and (d) $0.05$, green circles denote the rescaled results for varying stiffness with $\mathcal{S}=5.0/(\ell_0/\sigma)$, where the definition of $\phi$ is changed to $\phi\equiv N\pi d_{\rm ex}^2/4V$.}}}
	\label{Fig:S6}
\end{figure}

\begin{figure}[h]
	\centering
	\includegraphics[width=.95\textwidth]{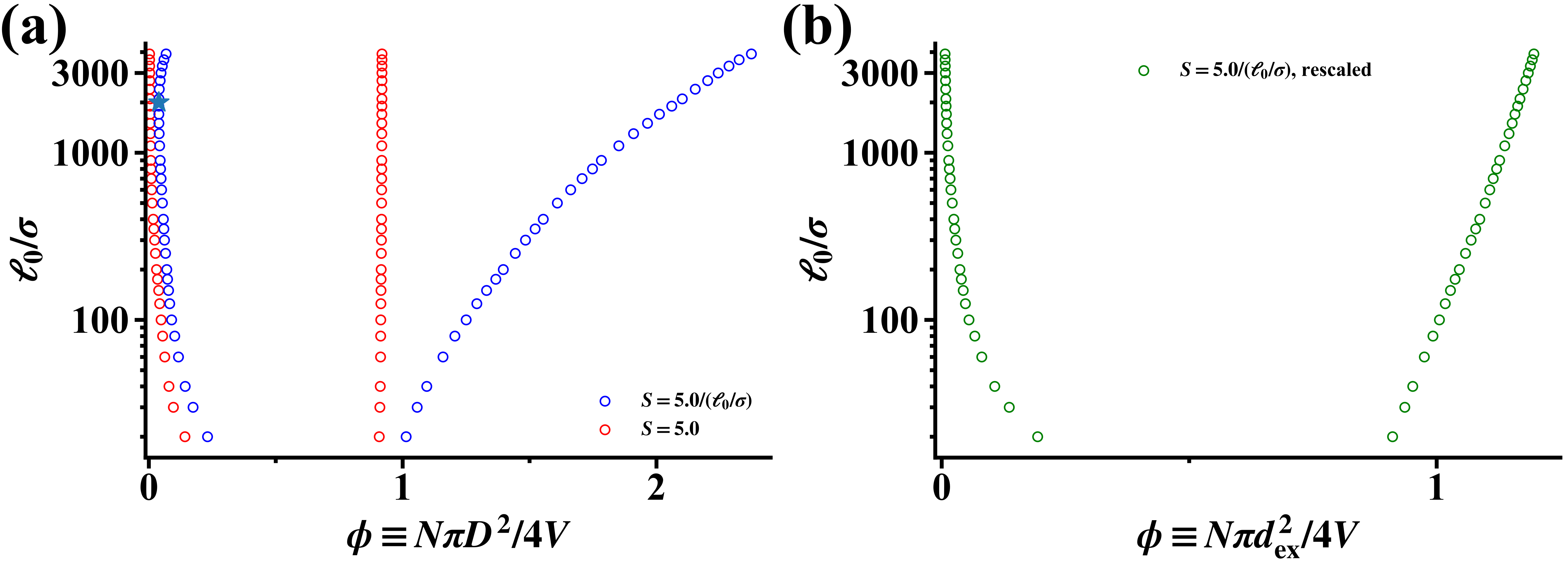}
	\caption{\protect\small{{Simulated binodal for 2D overdamped ABPs. In (a), blue circles denote varying stiffness with $\mathcal{S}=5.0/(\ell_0/\sigma)$, red circles denote fixed stiffness $\mathcal{S}=5.0$. The blue star denotes the critical activity where the gas phase density begins to increase with $\ell_0/\sigma$. In (b), green circles denote the rescaled results for varying stiffness with $\mathcal{S}=5.0/(\ell_0/\sigma)$, where the definition of $\phi$ is changed to $\phi\equiv N\pi d_{\rm ex}^2/4V$.}}}
	\label{Fig:S7}
\end{figure}

Figures~\ref{Fig:S6}(a) and (b) present our 2D simulation results for ${\rm St}=0.03$ and $0.05$.
When we allow for an activity dependent stiffness parameter of the form $\mathcal{S}=5.0/(\ell_0/\sigma)$, we observe a non-monotonic trend in the gas phase density (blue circles) with increasing activity.
Specifically, for ${\rm St}=0.03$, $\phi_{\rm gas}$ increases with $\ell_0/\sigma$ when $\ell_0/\sigma \gtrsim 1600$, while for ${\rm St}=0.05$, the transition begins at $\ell_0/\sigma \gtrsim 1000$.
Preparing a system at a volume fraction slightly larger than $\phi_{\rm gas}$ at these transition points would result in one observing MIPS. 
A slight increase in activity while keeping $\phi$ constant could then result in exiting the MIPS binodal as new gas phase density now exceeds $\phi$.
This is the so-called ``reentrance'' effect first reported by Ref.~\cite{Mandal2019Motility-InducedPhases}.
We also find that the critical activity for reentrance effect decreases with increasing inertia which is also in agreement with Ref.~\cite{Mandal2019Motility-InducedPhases}.
Fig.~\ref{Fig:S7}(a) presents our simulation results for 2D overdamped particles.
Crucially, when we allow for an activity dependent stiffness parameter, we again observe a non-monotonic trend in the gas phase density (blue circles) with increasing activity.
Specifically, $\phi_{\rm gas}$ increases with $\ell_0/\sigma$ when $\ell_0/\sigma \gtrsim 2000$, providing clear evidence that the ``softness'' of the interparticle interactions is responsible for the observed reentrance effect.

Before discussing the origins of the reentrance effect, it is notable that the activity required to see the non-monotonicity in $\phi_{\rm gas}$ is considerably higher in this worth than what was reported in Ref.~\cite{Mandal2019Motility-InducedPhases}.
We note that in Refs.~\cite{Mandal2019Motility-InducedPhases,Hecht22} the binodal was not directly computed. 
Rather, regions in parameter space were labeled as homogeneous or phase separated based on if MIPS was observed within some finite time duration. 
This can result in regions in which a homogeneous state is metastable (e.g., requires a rare event to phase separate) as being labeled as stable states when in fact MIPS is the preferred state. 
This, in turn, can lead one to conclude that the region of MIPS is narrower than in actuality. 
By directly measuring the binodal densities, our work avoids the possible issue.

\begin{figure}[h]
	\centering
	\includegraphics[width=.95\textwidth]{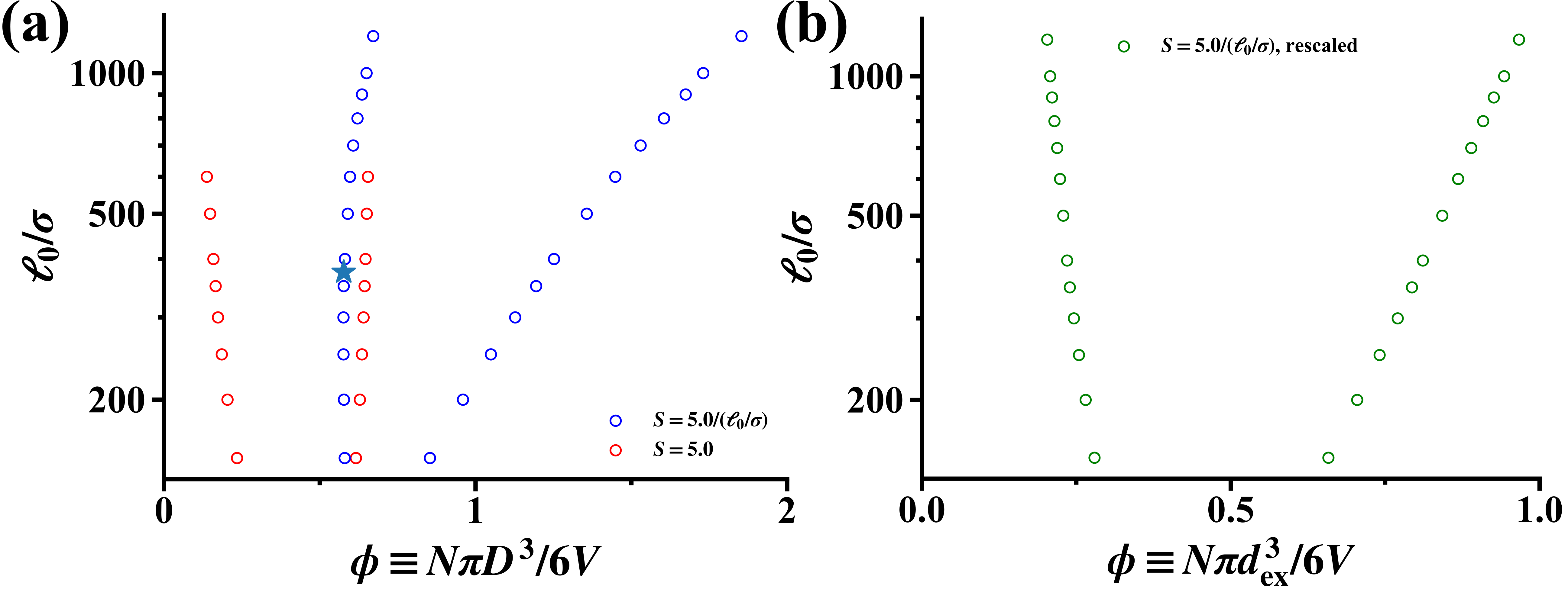}
	\caption{\protect\small{{Simulated binodal for 3D inertial ABPs for ${\rm St} = 0.04$. (a) Blue circles denote varying stiffness with $\mathcal{S} = 5.0/(\ell_0/\sigma)$, red circles denote fixed stiffness $\mathcal{S} = 5.0$. Blue star denotes the critical activity where $\phi_{\rm gas}$ begins to increase with activity. (b) Green circles denote rescaled results for $\mathcal{S} = 5.0/(\ell_0/\sigma)$, where the definition of $\phi$ is changed to $\phi\equiv N\pi d_{\rm ex}^3/6V$.}}}
	\label{Fig:S8}
\end{figure}

We now compute the binodal for a fixed stiffness with  $\mathcal{S}=5.0$, as shown by the red circles in Figs.~\ref{Fig:S6}(a)(b) and Fig.~\ref{Fig:S7}(b). 
The non-monotonic trend is absent for both values of inertia and overdamped particles with $\phi_{\rm gas}$ decreases continuously with increasing activity: \textit{variable stiffness (and not inertia) is the origin of the reentrance effect.}
We postulate that the central impact of the decreasing stiffness with activity is to reduce the effective diameter and, hence, the effective particle volume (area in 2D). 
In 2D (3D), $v_{\rm ex} = \pi d_{\rm ex}^2 / 4$ ($v_{\rm ex} = \pi d_{\rm ex}^3 / 6$), where $d_{\rm ex}$ is again determined through the radial distribution function. 
With decreasing stiffness, $d_{\rm ex}$ decreases as the increase in the relative active force allows for greater overlap. 
To examine if the changing $d_{\rm ex}$ accounts for the non-monotonic dependence of the gas phase density with activity, we redefine the volume fraction of the binodal curves using the variable effective diameter rather than the previously used diameter $D$ (see Figs.~\ref{Fig:S6}(c)(d) and Fig.~\ref{Fig:S7}(b)).
Accounting for the size variation introduced by the variable stiffness completely eliminates the non-monotonic trend in $\phi_{\rm gas}$.
The reentrance effect is not due to inertia but rather due to the variable particle stiffness.
As activities $\ell_0/\sigma$ can be as high as $\sim$3000 in Fig.~\ref{Fig:S6} and ~\ref{Fig:S7}, it is certainly true that the reentrance we reported in these figures could suffer from finite-size effects.
It is not obvious to us whether these effects would enhance or reduce the reentrant effect.
We note that the reentrance of MIPS for overdamped ABPs was also reported in Ref.~\cite{Bialke13}.
While inertia can influence the value of the volume exclusion diameter and therefore the location of the reentrance effect, it does not play a determining role in the effect.
The conclusions made here using our 2D simulation data can also be arrived at for a 3D system, as shown in Fig.~\ref{Fig:S8}.

\newpage

\newpage
\section{Fokker-Planck Analysis of Interacting Inertial ABPs}
Here we provide a systematic derivation of the coexistence criteria of MIPS for inertial ABPs that was summarized in the main text.
Starting from the Fokker-Planck equation describing the $N$-body probability distribution, we will obtain an expression for the dynamic stress which forms the basis of the nonequilibrium theory of coexistence for single-component systems~\cite{Omar23b}.
We note that while the equations for interacting ABPs have appeared through various forms in the literature~\cite{Paliwal18,Solon18,Epstein19, Omar23a,Omar23b}, additional closures and approximations will need to be introduced to allow us to determine the coexistence criteria for inertial active matter.

\subsection{Exact Formulations}
\label{sec:exact_formulations}
The particles obey the underdamped translational Langevin equation-of-motion:
\begin{subequations}
\begin{equation}
    m\dot{\mathbf{u}}{_\alpha} = \mathbf{F}^{\rm A}_{\alpha} - \zeta \mathbf{u}_{\alpha}+\mathbf{F}_{\alpha}^{\rm ext}+\mathbf{F}_{\alpha}^{\rm B}+\sum \limits_{\beta \neq \alpha} \mathbf{F}_{\alpha \beta}^{\rm C} \ ,
\label{eq:equation-of-motion}
\end{equation}
where the subscript $\alpha$ runs from $1$ to $N$. 
Note that in the remaining part of this manuscript, we will use $\mathbf{u}_{\alpha}$ in place of $\dot{\mathbf{x}}_{\alpha}$ to denote the velocity of particle $\alpha$ for notational convenience.
Every particle experiences five distinct forces: an active force $\mathbf{F}^{\rm A}_{\alpha}=\zeta U_0 \mathbf{q}_{\alpha}$ ($\zeta$ is the translational drag coefficient and $U_0$ is the intrinsic active speed), a drag force $\mathbf{F}^{\rm drag}_{\alpha}=- \zeta \mathbf{u}_{\alpha}$, an external force $\mathbf{F}_{\alpha}^{\rm ext}$, a stochastic Brownian force $\mathbf{F}_{\alpha}^{\rm B}$, and an interparticle (pairwise) conservative force $\mathbf{F}_{\alpha \beta}^{\rm C}$.
The Brownian forces satisfy fluctuation-dissipation theorem $\langle \mathbf{F}_{\alpha}^{\rm B}(t) \mathbf{F}_{\alpha}^{\rm B}(t^{'}) \rangle = 2D^{\rm B} \zeta^2 \delta(t-t^{'}) \mathbf{I} = 2k_{\rm B} T \zeta \delta(t-t^{'}) \mathbf{I}$.
The rotary dynamics are \textit{overdamped}:
\begin{equation}
\dot{\mathbf{q}}_{\alpha}=\mathbf{\Omega}_{\alpha}^{\rm R} \times \mathbf{q}_{\alpha} \ ,
\label{eq:rotary}
\end{equation}
\end{subequations}
where the stochastic angular velocity $\mathbf{\Omega}_{\alpha}^{\rm R}$ has statistics:
\begin{equation*}
    \langle \mathbf{\Omega}_{\alpha}^{\rm R} (t) \rangle =\mathbf{0}, \ \langle \mathbf{\Omega}_{\alpha}^{\rm R} (t) \mathbf{\Omega}_{\beta}^{\rm R} (t^{'}) \rangle = \frac{2}{\tau_{\rm R}} \delta_{\alpha \beta} \delta (t-t^{'}) \mathbf{I} \ .
\end{equation*}

The probability density of finding the system in a microstate $\mathbf{\Gamma}=(\mathbf{x}^N, \mathbf{u}^N, \mathbf{q}^N)$ at time $t$ is denoted as $f_N(\mathbf{\Gamma}; t)$. 
The Fokker-Planck equation governing this distribution follows from the equations-of-motion [Eq.~\eqref{eq:equation-of-motion}-~\eqref{eq:rotary}] with
\begin{equation}
\label{eq:F-P}
    \frac{\partial f^N(\mathbf{\Gamma}, t)}{\partial t}+\sum \limits_{\alpha} (\bm{\nabla}_{\alpha} \cdot \mathbf{j}^{\mathbf{x}}_{\alpha} + \bm{\nabla}_{\alpha}^{\mathbf{u}} \cdot \mathbf{j}^{\mathbf{u}}_{\alpha} + \bm{\nabla}_{\alpha}^{\mathbf{q}} \cdot \mathbf{j}^{\mathbf{q}}_{\alpha})=0 \ ,
\end{equation}
where
\begin{align*}
    \bm{\nabla}_{\alpha} =& \frac{\partial}{\partial \mathbf{x}_{\alpha}}, \ \bm{\nabla}_{\alpha}^{\mathbf{u}} = \frac{\partial}{\partial \mathbf{u}_{\alpha}}, \  \bm{\nabla}_{\alpha}^{\mathbf{q}} = \mathbf{q}_{\alpha} \times \frac{\partial}{\partial \mathbf{q}_{\alpha}} \ , \nonumber \\ 
    \mathbf{j}^{\mathbf{x}}_{\alpha} =& \mathbf{u}_{\alpha}f_N  \ , \\
    \mathbf{j}^{\mathbf{u}}_{\alpha} =& \frac{1}{m} (\zeta U_0 \mathbf{q}_{\alpha} f_N - \zeta \mathbf{u}_{\alpha}f_N + \mathbf{F}_{\alpha}^{\rm ext} f_N  +\sum \limits_{\beta \neq \alpha} \mathbf{F}_{\alpha \beta}^{\rm C}f_N - D_{\rm B} \zeta^2 \bm{\nabla}_{\alpha}^{\mathbf{u}} f_N ) \ , \\
    \mathbf{j}^{\mathbf{q}}_{\alpha} =& -\frac{1}{\tau_R} \bm{\nabla}_{\alpha}^{\mathbf{q}} f_N \ .
\end{align*}
It proves convenient to define the following Fokker-Planck operator:
\begin{equation*}
    \mathcal{L} = \sum \limits_{\alpha} \Big [ -\bm{\nabla}_{\alpha} \cdot \mathbf{u}_{\alpha} +\frac{1}{m} \bm{\nabla}_{\alpha}^{\mathbf{u}} \cdot \Big (-\zeta U_0 \mathbf{q}_{\alpha} + \zeta \mathbf{u}_{\alpha} -\mathbf{F}_{\alpha}^{\rm ext}    - \sum \limits_{\beta \neq \alpha} \mathbf{F}_{\alpha \beta}^{\rm C}   + D_{\rm B} \zeta^2  \bm{\nabla}_{\alpha}^{\mathbf{u}} \Big ) +  \bm{\nabla}_{\alpha}^{\mathbf{q}} \cdot \Big (\frac{1}{\tau_{\rm R}}\bm{\nabla}_{\alpha}^{\mathbf{q}}\Big ) \Big ] \ ,
\end{equation*} 
such that Eq.~\eqref{eq:F-P} can be compactly expressed as $\partial f/\partial t = \mathcal{L}f$.

A solution for the full $N$-body distribution function $f_N$ allows for the determination of the statistics of any observable $\hat{\mathcal{O}}$, including the ensemble average with $\mathcal{O}=\langle \hat{\mathcal{O}} (\mathbf{\Gamma}) \rangle$, where $\langle \cdot \rangle \equiv \int ( \cdot ) f_N (\mathbf{\Gamma}, t) d \mathbf{\Gamma}$ denotes ensemble average and $\hat{\mathcal{O}} (\mathbf{\Gamma})$ is the microscopic definition of the observable.
The evolution of the average of the observable follows as
\begin{equation}
    \label{eq:evolution}
    \frac{\partial \mathcal{O}}{\partial t} = \int \hat{\mathcal{O}} \left ( \frac{\partial f_N}{\partial t} \right ) d \mathbf{\Gamma} = \int \hat{\mathcal{O}}( \mathcal{L} f_N ) d\mathbf{\Gamma} = \int (\mathcal{L}^{\dagger} \hat{\mathcal{O}} ) f_N d\mathbf{\Gamma} \ ,
\end{equation}
where the adjoint of $\mathcal{L}$ can be expressed as
\begin{equation}
    \mathcal{L}^{\dagger} = \sum \limits_{\alpha} \Big [ \mathbf{u}_{\alpha} \cdot \bm{\nabla}_{\alpha} + \frac{1}{m} \Big (\zeta U_0 \mathbf{q}_{\alpha} - \zeta \mathbf{u}_{\alpha} +\mathbf{F}_{\alpha}^{\rm ext} + \sum \limits_{\beta \neq \alpha} \mathbf{F}_{\alpha \beta}^{\rm C} + D_{\rm B} \zeta^2  \bm{\nabla}_{\alpha}^{\mathbf{u}} \Big ) \cdot \bm{\nabla}_{\alpha}^{\mathbf{u}} + \frac{1}{\tau_{\rm R}}   \bm{\nabla}_{\alpha}^{\mathbf{q}} \cdot \bm{\nabla}_{\alpha}^{\mathbf{q}} \Big ] \ .
    \label{eq:adjoint}
\end{equation}

Eq.~\eqref{eq:evolution} and ~\eqref{eq:adjoint} allow us to obtain the evolution equations for any observable. 
In the case of phase separation, the order parameter that distinguishes the two phases is the (number) density. 
We thus derive the evolution equation for density field (defined microscopically as $\hat{\rho} (\mathbf{x}) = \sum_{\alpha} \delta (\mathbf{x}- \mathbf{x}_{\alpha} )$) with: 
\begin{equation}
\label{eq:rho}
    \frac{\partial \rho}{\partial t}=- \bm{\nabla} \cdot \mathbf{j}^{\rho} \ ,
\end{equation}
where $\hat{\mathbf{j}}^{\rho}(\mathbf{x})= \sum_{\alpha} \mathbf{u}_{\alpha} \delta (\mathbf{x}- \mathbf{x}_{\alpha} )$ is the number density flux.
We now require the evolution  equation for the number density flux:
\begin{equation}
\label{eq:jrho}
    m\frac{\partial \mathbf{j}^{\rho}}{\partial t}=\bm{\nabla} \cdot \underbrace{( \bm{\sigma}^{\rm K} + \bm{\sigma}^{\rm C} )}_{{\rm stresses}} + \underbrace{\zeta U_0 \mathbf{m} - \zeta \mathbf{j}^{\rho} + \mathbf{F}^{\rm ext} \rho }_{{\rm body \ forces}}  \ ,
\end{equation}
where $\bm{\sigma}^{\rm K}$, $\bm{\sigma}^{\rm C}$, and $\mathbf{m}$ are kinetic stress, interaction stress, and polarization density, respectively.
Their microscopic definitions follow as:
\begin{align}
 \mathbf{m} &= \big\langle \sum_{\alpha} \mathbf{q}_{\alpha} \delta (\mathbf{x} - \mathbf{x}_{\alpha} ) \big\rangle \\
    \bm{\sigma}^{\rm K} &= \big\langle -m \sum \limits _{\alpha} \mathbf{u}_{\alpha} \mathbf{u}_{\alpha} \delta (\mathbf{x}- \mathbf{x}_{\alpha} ) \big\rangle \ , \\
    \bm{\sigma}^{\rm C} &= \big\langle -\frac{1}{2} \sum \limits _{\alpha} \sum \limits _{\beta \neq \alpha} \mathbf{x}_{\alpha \beta} \mathbf{F}_{\alpha \beta}^{\rm C} b_{\alpha \beta} \big\rangle \ ,
\end{align}
where the bond function is defined as $b_{\alpha \beta} = \int_0^1 d\lambda \delta ( \lambda \mathbf{x}_{\alpha \beta} + \mathbf{x}_{\beta} - \mathbf{x})$ and $\mathbf{x}_{\alpha \beta}= \mathbf{x}_{\alpha}- \mathbf{x}_{\beta}$~\cite{Hardy82}.

Equation~\eqref{eq:jrho} is nothing more than a linear momentum balance and highlights that stresses and body forces determine the spatial and temporal evolution of the particle flux. 
To describe the particle flux (and, consequently, the density field dynamics) we require expressions for both the stresses and the polarization density.
Beginning with the polarization density we find: 
\begin{equation}
\label{eq:m}
    \frac{\partial \mathbf{m}}{\partial t } + \frac{d-1}{\tau_{\rm R}} \mathbf{m} + \bm{\nabla} \cdot \mathbf{j}^{\mathbf{m}} = 0 \ ,
\end{equation}
where $ \hat{\mathbf{j}}^{\mathbf{m}} =  \sum _{\alpha}\mathbf{u}_{\alpha} \mathbf{q}_{\alpha} \delta (\mathbf{x} - \mathbf{x}_{\alpha} )  $ is the polarization flux given by:
\begin{equation}
    m \frac{\partial \mathbf{j}^{\mathbf{m}}}{\partial t} = \bm{\nabla}\cdot \mathbf{D}^{\mathbf{m}} + \zeta U_0 \Tilde{\mathbf{Q}} - \zeta  \mathbf{j}^{\mathbf{m}} + \mathbf{F}^{\rm ext} \mathbf{m} + \bm{\kappa}^{\mathbf{m}} +\bm{\nabla} \cdot \bm{\Sigma}^{\mathbf{m}} -m \frac{d-1}{\tau_{\rm R}} \mathbf{j}^{\mathbf{m}} \ .
    \label{eq:jm}
\end{equation}
The polarization flux dynamics contain a convective contribution proportional to the nematic order density $\Tilde{\mathbf{Q}}$:
\begin{equation}
    \Tilde{\mathbf{Q}} = \big \langle \sum\limits_{\alpha} \mathbf{q}_{\alpha} \mathbf{q}_{\alpha} \delta(\mathbf{x}- \mathbf{x}_{\alpha}) \big \rangle \ , 
\end{equation}
and additional contributions from the following terms:
\begin{align}
    \mathbf{D}^{\mathbf{m}} &= \big \langle -m \sum\limits_{\alpha} \mathbf{u}_{\alpha}   \mathbf{u}_{\alpha} \mathbf{q}_{\alpha} \delta(\mathbf{x}- \mathbf{x}_{\alpha} )\big \rangle  \ , \\
    \bm{\kappa}^{\mathbf{m}} &= \big \langle \frac{1}{2} \sum \limits _{\alpha} \sum \limits _{\beta \neq \alpha} \mathbf{F}_{\alpha \beta}^{\rm C} \mathbf{q}_{\alpha \beta} b_{\alpha \beta} \big \rangle  \ , \\
    \bm{\Sigma}^{\mathbf{m}} &= \big \langle -\frac{1}{2} \sum \limits _{\alpha} \sum \limits _{\beta \neq \alpha} \mathbf{x}_{\alpha \beta} \mathbf{F}_{\alpha \beta}^{\rm C} \mathbf{d}_{\alpha \beta} \big \rangle \ ,
\end{align}
where $\mathbf{d}_{\alpha \beta}=\int_0^1 d\lambda (\mathbf{q}_{\beta}+\lambda \mathbf{q}_{\alpha \beta} ) \delta (\mathbf{x} - \mathbf{x}_{\beta}- \lambda \mathbf{x}_{\alpha \beta} )$, $\mathbf{q}_{\alpha \beta}=\mathbf{q}_{\alpha}-\mathbf{q}_{\beta}$.
The generalized Irving-Kirkwood-Noll procedure (see Refs.~\cite{Irving50,Lehoucq10,Langford24} for additional details) was used to derive the forms of $\bm{\kappa}^{\mathbf{m}}$ and $\bm{\Sigma}^{\mathbf{m}}$.
In principle, expressions for $\mathbf{j}^{\mathbf{m}}$, $\Tilde{\mathbf{Q}}$, $\mathbf{D}^{\mathbf{m}}$, $\bm{\kappa}^{\mathbf{m}}$, and $\bm{\Sigma}^{\mathbf{m}}$ are required to describe the polarization dynamics. 
Later, following Ref.~\cite{Omar23b}, we will propose a constitutive equation for $\bm{\kappa}^{\mathbf{m}}$ and neglect the quantitative impact of $\bm{\Sigma}^{\mathbf{m}}$.

The dynamics of the nematic field are given by:
\begin{equation}
\label{eq:Q}
    \frac{\partial \Tilde{\mathbf{Q}}}{\partial t} + \frac{2d}{\tau_{\rm R}} ( \Tilde{\mathbf{Q}} - \frac{1}{d}\rho \mathbf{I}) + \bm{\nabla} \cdot \mathbf{j}^{\Tilde{\mathbf{Q}}} = 0 \ ,
\end{equation}
where the nematic flux $\mathbf{j}^{\Tilde{\mathbf{Q}}}$ is defined as
\begin{equation}
    \mathbf{j}^{\Tilde{\mathbf{Q}}} = \big \langle \sum \limits _{\alpha} \mathbf{u}_{\alpha} \mathbf{q}_{\alpha} \mathbf{q}_{\alpha} \delta(\mathbf{x}- \mathbf{x}_{\alpha})  \big \rangle 
\end{equation}
and satisfies the following evolution equation: 
\begin{equation}
\label{eq:jQ}
    m \frac{\partial j^{\Tilde{\mathbf{Q}}}_{ijk}}{\partial t} = \partial^l G_{lijk} + \zeta U_0 \Tilde{B}_{ijk} - \zeta j^{\Tilde{\mathbf{Q}}}_{ijk} + m \frac{2}{\tau_{\rm R}} \delta_{jk} j^{\rho}_i + F^{\rm ext}_i \Tilde{Q}_{jk} + \kappa^{\Tilde{\mathbf{Q}}}_{ijk} + \partial^l \Sigma^{\Tilde{\mathbf{Q}}}_{lijk}-m \frac{2d}{\tau_{\rm R}} j^{\Tilde{\mathbf{Q}}}_{ijk} \ ,
\end{equation}
where we have invoked indicial notation.
The nematic field dynamics contain a convective contribution proportional to the one-body third orientational moment:
\begin{equation}
    \Tilde{B}_{ijk} = \big \langle \sum \limits _{\alpha} q_{\alpha,i} q_{\alpha,j} q_{\alpha,k } \delta(\mathbf{x}-\mathbf{x}_{\alpha} ) \big \rangle \ ,
\end{equation}
and additional contributions from the following terms:
\begin{align}
    G_{lijk} &= \big \langle -m \sum \limits _{\alpha} u_{\alpha,l} u_{\alpha,i} q_{\alpha,j} q_{\alpha,k} \delta(\mathbf{x}-\mathbf{x}_{\alpha} ) \big \rangle  \ , \\
    \kappa^{\Tilde{\mathbf{Q}}}_{ijk} &= \big \langle \frac{1}{2} \sum \limits _{\alpha} \sum \limits_{\beta \neq \alpha} F_{\alpha \beta, i}^{\rm C} (q_{\alpha, j} q_{\alpha, k} - q_{\beta, j} q_{\beta, k}) b_{\alpha \beta} \big \rangle \ , \\
    \Sigma^{\Tilde{\mathbf{Q}}}_{lijk} &= \big \langle -\frac{1}{2} \sum \limits _{\alpha} \sum \limits_{\beta \neq \alpha} x_{\alpha \beta,l} F_{\alpha \beta, i}^{\rm C} f_{\alpha \beta, jk} \big \rangle \ ,
\end{align} 
where $\mathbf{f}_{\alpha \beta}=\int_0^1 d\lambda (\mathbf{q}_{\beta}\mathbf{q}_{\beta}+\lambda (\mathbf{q}_{\alpha}\mathbf{q}_{\alpha} - \mathbf{q}_{\beta}\mathbf{q}_{\beta} )) \delta (\mathbf{x} - \mathbf{x}_{\beta}- \lambda \mathbf{x}_{\alpha \beta} )$.
In the above notation, $u_{\alpha,l}$ indicates the $l$th component of $\mathbf{u}_{\alpha}$.

The dynamics of $\mathbf{D}^{\mathbf{m}}$ are:
\begin{align}
\label{eq:Dm}
    \frac{\partial D^{\mathbf{m}}_{ijk}}{\partial t} = &-\partial^l E_{lijk} - \zeta U_0 (j^{\Tilde{\mathbf{Q}}}_{ijk}+j^{\Tilde{\mathbf{Q}}}_{jik})  - 2D_{\rm B} \zeta^2 \delta_{ij} m_k - (F_i^{\rm ext} j_{jk}^{\mathbf{m}} + F_j^{\rm ext} j_{ik}^{\mathbf{m}} ) - \partial^l ( \Sigma^{\mathbf{j}^{\mathbf{m}}}_{lijk}+\Sigma^{\mathbf{j}^{\mathbf{m}}}_{ljik})  \nonumber \\ & - (\kappa^{\mathbf{j}^{\mathbf{m}}}_{ijk} + \kappa^{\mathbf{j}^{\mathbf{m}}}_{jik}) -\frac{2\zeta}{m} D^{\mathbf{m}}_{ijk} - \frac{d-1}{\tau_{\rm R}} D^{\mathbf{m}}_{ijk} \ ,
\end{align} 
where
\begin{align}
    E_{lijk} &= \big \langle -m \sum \limits _{\alpha} u_{\alpha,l} u_{\alpha,i} u_{\alpha,j} q_{\alpha,k} \delta(\mathbf{x}-\mathbf{x}_{\alpha} ) \big \rangle \ , \\
    \kappa^{\mathbf{j}^{\mathbf{m}}}_{ijk} &= \big \langle \frac{1}{2} \sum \limits _{\alpha} \sum \limits_{\beta \neq \alpha} F_{\alpha \beta, i}^{\rm C} (u_{\alpha, j} q_{\alpha, k} - u_{\beta, j} q_{\beta, k}) b_{\alpha \beta} \big \rangle \ , \\
    \Sigma^{\mathbf{j}^{\mathbf{m}}}_{lijk} &= \big \langle -\frac{1}{2} \sum \limits _{\alpha} \sum \limits_{\beta \neq \alpha} x_{\alpha \beta,l} F_{\alpha \beta, i}^{\rm C} e_{\alpha \beta, jk} \big \rangle \ ,
\end{align}
and $\mathbf{e}_{\alpha \beta}=\int_0^1 d\lambda (\mathbf{u}_{\beta}\mathbf{q}_{\beta}+\lambda (\mathbf{u}_{\alpha}\mathbf{q}_{\alpha} - \mathbf{u}_{\beta}\mathbf{q}_{\beta} )) \delta (\mathbf{x} - \mathbf{x}_{\beta}- \lambda \mathbf{x}_{\alpha \beta} )$. 

The evolution equation for the kinetic stress $\bm{\sigma}^{\rm K}$ is:
\begin{align}
    \frac{\partial \bm{\sigma}^{\rm K}}{\partial t} &= - \bm{\nabla} \cdot \bm{\sigma}^{\rm K} - \zeta U_0 (\mathbf{j}^{\mathbf{m}} + (\mathbf{j}^{\mathbf{m}})^{\rm T} ) - \frac{2\zeta }{m} \bm{\sigma}^{\rm K} - (\mathbf{F}^{\rm ext} \mathbf{j}^{\rho} + \mathbf{j}^{\rho} \mathbf{F}^{\rm ext}) - (\bm{\kappa}^{\mathbf{j}^{\rho}} + (\bm{\kappa}^{\mathbf{j}^{\rho}})^{\rm T} )  \nonumber \\ &- \partial^k ( \Sigma^{\mathbf{j}^{\rho}}_{kij} + \Sigma^{\mathbf{j}^{\rho}}_{kji}) - 2D_{\rm B} \zeta^2 \rho \mathbf{I} \ .
\label{eq:sigmaK}
\end{align}
The additional tensors introduced in Eq.~\eqref{eq:sigmaK} take the following form:
\begin{align}
    \bm{\Sigma}^{\rm K} &= \big \langle -m \sum \limits_{\alpha} \mathbf{u}_{\alpha} \mathbf{u}_{\alpha} \mathbf{u}_{\alpha} \delta (\mathbf{x} - \mathbf{x}_{\alpha}) \big \rangle  \ ,\\
    \bm{\kappa}^{\mathbf{j}^{\rho}} &= \big \langle \frac{1}{2} \sum \limits _{\alpha} \sum \limits _{\beta \neq \alpha} \mathbf{F}_{\alpha \beta}^C \mathbf{u}_{\alpha \beta} b_{\alpha \beta} \big \rangle \ , \\
    \bm{\Sigma}^{\mathbf{j}^{\rho}} &=  \big \langle -\frac{1}{2} \sum \limits _{\alpha} \sum \limits _{\beta \neq \alpha} \mathbf{x}_{\alpha \beta} \mathbf{F}_{\alpha \beta}^C \mathbf{c}_{\alpha \beta} \big \rangle \ ,
\end{align}
where $\mathbf{c}_{\alpha \beta}=\int_0^1 d\lambda (\mathbf{q}_{\beta}+\lambda \mathbf{q}_{\alpha \beta} ) \delta (\mathbf{r} - \mathbf{r}_{\beta}- \lambda \mathbf{r}_{\alpha \beta} )$.

\subsection{Constitutive Equations}
Thus far, the equations provided are exact but now require constitutive equations, closures and other approximations that are appropriate for our context of describing states of phase coexistence. 
Here, we introduce constitutive equations motivated by the microscopic expressions of $\bm{\kappa}^{\mathbf{m}}$, $\bm{\kappa}^{\mathbf{j}^{\mathbf{m}}}$, and $\bm{\kappa}^{\Tilde{\mathbf{Q}}}$.
Following Ref.~\cite{Omar23b}, we observe that configurations in which particles point in the same direction $\mathbf{q}_{\alpha}=\mathbf{q}_{\beta}$ do not contribute to $\bm{\kappa}^{\mathbf{m}}$ while configurations with antialigned particles $\mathbf{q}_{\alpha}=-\mathbf{q}_{\beta}$ contribute the most to $\bm{\kappa}^{\mathbf{m}}$ in magnitude.
We therefore identify that $\bm{\kappa}^{\mathbf{m}}$ is correlated with the reduction in the effective active speed $U_{\rm eff}$ due to interparticle interactions.
That is, a pair of particles slow down when they collide head to head but active motion is nearly unaffected when particles are oriented in the same direction.
This motivates the following constitutive equations
\begin{align}
\label{eq:kappam}
    \bm{\kappa}^{\mathbf{m}} &= -\zeta (U_0 - U_{\rm eff}^{\mathbf{m}} ) \Tilde{\mathbf{Q}} \ , \\
\label{eq:kappajm}
    \bm{\kappa}^{\mathbf{j}^{\mathbf{m}}} &= -\zeta (U_0 - U_{\rm eff}^{\mathbf{j}^{\mathbf{m}}}) \mathbf{j}^{\Tilde{\mathbf{Q}}} \ , \\
\label{eq:kappaQ}
    \bm{\kappa}^{\Tilde{\mathbf{Q}}} &= -\zeta (U_0 - U_{\rm eff}^{\Tilde{\mathbf{Q}}} ) \Tilde{\mathbf{B}} \ ,
\end{align}
where $U_{\rm eff}^{\mathbf{m}}=U_0 \overline{U}^{\mathbf{m}}$, $U_{\rm eff}^{\mathbf{j}^{\mathbf{m}}} = U_0 \overline{U}^{\mathbf{j}^{\mathbf{m}}}$, and $U_{\rm eff}^{\Tilde{\mathbf{Q}}}=U_0 \overline{U}^{\Tilde{\mathbf{Q}}}$ are the effective speed of polarization density transport, polarization flux transport and nematic order convection, respectively.
$\overline{U}^{\alpha} \in [ 0, 1 ]$ ($\alpha = \mathbf{m}, \mathbf{j}^{\mathbf{m}}, \Tilde{\mathbf{Q}}$) is defined as the corresponding dimensionless active speed.

\subsection{General Formulation of Coexistence Criteria}
We formulate our coexistence criteria by considering a quasi-1d phase separated system with a planar interface. 
Without loss of generality, we take $z$ to be the direction normal to the interface.
This geometry reduces the equations in the previous section to scalar equations and, as we are interested in \textit{stationary} phase coexistence, we can neglect all time derivatives in the previous section.
Consistent with these conditions, Ref.~\cite{Omar23b} found that the mechanical condition for a one-dimension stationary state (such as phase separation) is a uniform \textit{dynamic stress}, $\Sigma$, where $\Sigma$ is defined directly from the $z$-component of the static linear momentum balance, $d\sigma/dz + b \equiv d\Sigma/dz$, where $\sigma$ and $b$ are the true stresses and body forces, respectively.
The dynamic stress coincides with the true stress only for passive systems with no external forces present.
However, for nonequilibrium systems they are generally distinct.

The condition of uniform dynamic stress can be combined with a constitutive equation to arrive at the following:
\begin{equation}
\label{eq:sigma_expansion}
    -\Sigma=\mathcal{P}(\rho)-a(\rho)\frac{d^2 \rho}{dz^2}-b(\rho) \left (\frac{d\rho}{dz} \right )^2 = C \ ,
\end{equation}
where $\mathcal{P}(\rho)$ is the bulk dynamic pressure and $C$ is a constant.
Recognizing that the gradients vanish in the bulk phases, we arrive at the first coexistence criterion:
\begin{equation*}
    \mathcal{P}(\rho_{\rm liq}) = \mathcal{P}(\rho_{\rm gas}) = C = \mathcal{P}^{\rm coexist} \ .
\end{equation*}

In order to find the second criterion we multiply the gradient terms with a weighting function and integrate the result across the interface.
It is found that the weighting function that leads to the vanishing of the integral containing the interfacial gradient terms takes the following form~\cite{Aifantis83}:
\begin{equation*}
    E(\rho) = \frac{1}{a(\rho)}\exp \left ( 2\frac{b(\rho)}{a(\rho)} d\rho \right ) \ .
\end{equation*}
The second criterion is therefore
\begin{equation*}
    \int_{\rho_{\rm gas}}^{\rho_{\rm liq}} \left [ \mathcal{P}(\rho) - \mathcal{P}^{\rm coexist}\right  ] E(\rho) d\rho = 0 \ .
\end{equation*}
Our task is to now identify the form of the dynamic stress [and see if the form coincides with that proposed in Eq.~\eqref{eq:sigma_expansion}] by simplifying our exact equations, utilizing our constitutive equations, and making additional closures and approximations.

\subsection{Closures and Approximations}
Though nonequilibrium steady states may admit non-zero fluxes with interfaces of finite curvature~\cite{Tjhung18}, a phase-separated system with a planar interface will result in $\mathbf{j}^{\rho}=0$.
Eq.~\eqref{eq:jrho} is thus reduced to
\begin{equation}
\label{eq:mechbala}
    \mathbf{0}=\bm{\nabla} \cdot \bm{\sigma}^{\rm K} +\zeta U_0 \mathbf{m}  + \bm{\nabla} \cdot \bm{\sigma}^{\rm C} \ .
\end{equation}
From Eq.~\eqref{eq:m} we can get an expression for the polarization density field
\begin{equation}
\label{eq:polafield}
    \mathbf{m}= - \frac{\tau_R}{d-1} \bm{\nabla}\cdot \mathbf{j}^{\mathbf{m}} \ .
\end{equation}
Substitution of Eq.~\eqref{eq:polafield} into Eq.~\eqref{eq:mechbala} allows us to express the dynamic stress as:
\begin{equation}
\label{eq:dynstresscontributions}
    \bm{\Sigma}= \bm{\sigma}^{\rm K} + \bm{\sigma}^{\rm act} +\bm{\sigma}^{\rm C} \ ,
\end{equation}
where we have defined the active stress:
\begin{equation}
\label{eq:sigmaact}
    \bm{\sigma}^{\rm act} = -\zeta U_0\frac{ \tau_{\rm R}}{d-1} \mathbf{j}^{\mathbf{m}} \ .
\end{equation}

We now aim to express $\Sigma=\sigma^{\rm K}_{zz}+\sigma^{\rm act}_{zz}+\sigma^{\rm C}_{zz}$ in terms of bulk equations of state and gradient expansions in density up to second order, consistent with Eq.~\eqref{eq:sigma_expansion}.
This will allow us to directly use the coexistence criteria found in Ref.~\cite{Omar23b}.
A gradient expansion of the conservative interparticle stress $\sigma^{\rm C}_{zz}$ results in the bulk interparticle pressure $p_{\rm int}(\rho)$ and Korteweg-like terms due to the distortion of the pair-distribution function in the presence of density gradients.
It can be shown that the coefficients on the gradient terms associated with $\sigma^{\rm C}_{zz}$ scale as $\zeta U_0 D$~\cite{Omar20}, while the gradient terms in the active stress scale as $\zeta U_0 \ell_0$. 
As MIPS occurs at $\ell_0/D \gg 1$, we can discard the Korteweg-like terms and approximate the conservative interparticle stress as $\sigma^{\rm C}_{zz} \approx -p_{\rm int}$.

For uniform and stationary systems, energy conservation requires $p_{\rm k}/p_{\rm act}=(d-1){\rm St}$.
We recognize that we can recover this result from Eq.~\eqref{eq:sigmaK} when $\bm{\nabla} \cdot \bm{\sigma}^{\rm K} + (\bm{\kappa}^{\mathbf{j}^{\rho}} + (\bm{\kappa}^{\mathbf{j}^{\rho}})^{\rm T} )+ \partial^k ( \Sigma^{\mathbf{j}^{\rho}}_{kij} + \Sigma^{\mathbf{j}^{\rho}}_{kji}) =0 $, which leads to:
\begin{equation}
\label{eq:k_act_rel}
    \frac{d-1}{\tau_{\rm R}} (\bm{\sigma}^{\rm act}+(\bm{\sigma}^{\rm act})^{\rm T})=\frac{2\zeta}{m} \bm{\sigma}^{\rm K} \ .
\end{equation}
Focusing on the relevant component of the stress tensor for 1D phase separation, we arrive at: 
\begin{equation}
\label{eq:propor}
    \sigma^{\rm K}_{zz}=(d-1){\rm St} \sigma^{\rm act}_{zz} \ .
\end{equation}
While Eq.~\eqref{eq:propor} is only strictly true for stationary isotropic systems, we \textit{postulate} that it holds locally despite the presence of an inhomogeneous density profile.
The result of this postulate is that the dynamic stress can now be expressed as
\begin{equation}
\label{eq:sigmaprimary}
    -\Sigma = p_{\rm int} - (1+(d-1){\rm St}) \sigma_{zz}^{\rm act}  =  p_{\rm int} + \zeta U_0 \tau_{\rm R} \frac{1+(d-1){\rm St}  }{d-1}j^{\mathbf{m}}_{zz} \ ,
\end{equation}
where a constitutive equation for $j^{\mathbf{m}}_{zz}$ is all that is required.

We now simplify the evolution equation for $j^{\mathbf{m}}_{zz}$ [Eq.~\eqref{eq:jm}] by defining the traceless nematic order  $\mathbf{Q}=\Tilde{\mathbf{Q}}- \rho \mathbf{I}/d$ and neglecting $\bm{\Sigma}^{\mathbf{m}}$ in Eq.~\eqref{eq:jm} (justified numerically in Ref.~\cite{Omar23b}):
\begin{equation}
\label{eq:oneplusjm}
    (1+(d-1){\rm St}) j^{\mathbf{m}}_{zz} = \frac{1}{\zeta} \frac{d}{dz} D^{\mathbf{m}}_{zzz} + U_0 \overline{U}^{\mathbf{m}} Q_{zz} + \frac{1}{d} U_0 \overline{U}^{\mathbf{m}}  \rho \ .
\end{equation}
We can simplify the nematic order evolution equation [Eq.~\eqref{eq:Q}] to find:
\begin{equation}
    Q_{zz} = - \frac{\tau_{\rm R}}{2d} \frac{d}{dz} j^{\Tilde{\mathbf{Q}}}_{zzz} \ .
\end{equation}
Before describing $j^{\Tilde{\mathbf{Q}}}_{zzz}$ and $D^{\mathbf{m}}_{zzz}$, we first define the traceless third orientational moment as $\mathbf{B}=\Tilde{\mathbf{B}}-\bm{\alpha} \cdot \mathbf{m} / (d+2)$, where $\alpha_{ijkl}=\delta_{ij} \delta_{kl} + \delta_{ik} \delta_{jl} + \delta_{il} \delta_{jk}$ is a fourth-rank isotropic tensor.
The evolution equations for $j^{\Tilde{\mathbf{Q}}}_{zzz}$ and $D^{\mathbf{m}}_{zzz}$ now simplify to: 
\begin{equation}
    (1+2d {\rm St}) j^{\Tilde{\mathbf{Q}}}_{zzz} = \frac{3}{d+2} U_0 \overline{U}^{\Tilde{\mathbf{Q}}} m_z \ ,
\end{equation}
and 
\begin{equation}
\label{eq:Dmzzz}
    (2+(d-1){\rm St}) D^{\mathbf{m}}_{zzz} = -2mU_0 \overline{U}^{\mathbf{j}^{\mathbf{m}}} j^{\Tilde{\mathbf{Q}}}_{zzz} \ ,
\end{equation}
where we have discarded terms arising from $\mathbf{G}$, $\mathbf{B}$, $\bm{\Sigma}^{\Tilde{\mathbf{Q}}}$, $\mathbf{E}$ and $\bm{\Sigma}^{\mathbf{j}^{\mathbf{m}}}$ as these will all contribute to higher order in spatial density gradients than what our coexistence theory permits.

We finally approximate that different effective speeds of active transport are identical, with $\overline{U}(\ell_0/\sigma, \phi,$ ${\rm St}) = \overline{U}^{\mathbf{m}} = \overline{U}^{\Tilde{\mathbf{Q}}} = \overline{U}^{\mathbf{j}^{\mathbf{m}}}$. 
Substitution of Eqs.~\eqref{eq:oneplusjm}-\eqref{eq:Dmzzz} into Eq.~\eqref{eq:sigmaprimary} we find:
\begin{equation}
\label{eq:sigmapri}
    -\Sigma =  \ p_{\rm int} + p_{\rm act} + p_{\rm k} - \frac{3\zeta \ell_0^2 U_0 }{2d(d-1)(d+2)(1+2d{\rm St})}  \Big[  \frac{4d {\rm St}}{ 2+(d-1){\rm St}} \frac{d}{dz} \big( \overline{U}^2 m_z  \big) 
     +  \overline{U} \frac{d}{dz} \big(\overline{U} m_z \big) \Big]   \ ,
\end{equation}
where we now define the bulk contributions to the active and kinetic stresses as:
\begin{equation}
  p_{\rm act} + p_{\rm k}=  \frac{\rho \zeta \ell_0 U_0 \overline{U}} {d(d-1)} \ .
  \label{eq:pact}
\end{equation}
Note that with $p_{\rm k}/p_{\rm act}=(d-1){\rm St}$, $p_{\rm k}$ and $p_{\rm act}$ can be isolated. 

We can eliminate the polarization from the dynamic stress by combining Eqs.~\eqref{eq:mechbala},~\eqref{eq:polafield},~\eqref{eq:sigmaact}, and ~\eqref{eq:propor}:
\begin{equation}
\label{eq:sumathbfz}
(1+(d-1){\rm St})\zeta U_0 m_z =  \frac{d p_{\rm int}}{dz} \ .
\end{equation}
We substitute Eq.~\eqref{eq:sumathbfz} into Eq.~\eqref{eq:sigmapri} and finally arrive at:
\begin{align}
\label{eq:sigmafin}
    -\Sigma &=  \ p_{\rm int} + p_{\rm act} + p_{\rm k}- \frac{3 \ell_0^2  }{2d(d-1)(d+2)(1+2d{\rm St})(1+(d-1){\rm St})}   \nonumber \\
     &\Big[  \frac{4d {\rm St}}{ 2+(d-1){\rm St}} \frac{d}{dz} \big( \overline{U}^2 \frac{d p_{\rm int}}{dz}  \big) +  \overline{U} \frac{d}{dz} \big(\overline{U} \frac{d p_{\rm int}}{dz} \big) \Big]   \ .
\end{align}

\subsection{Coexistence Criteria of Inertial ABPs}
With our dynamic stress now precisely in the form needed [Eq.~\eqref{eq:sigmafin}] to apply the mechanical theory of nonequilibrium coexistence [Eq.~\eqref{eq:sigma_expansion}], we can identify: 
\begin{subequations}
\begin{align}
    \mathcal{P}(\rho) =& \  p_{\rm int}+p_{\rm act}+p_{\rm  k}  \ , \label{eq:P} \\
    a(\rho) =&   \frac{3 \ell_0^2 }{2d(d-1)(d+2)(1+2d{\rm St})(1+(d-1){\rm St})} \Big[ \frac{4d {\rm St}}{ 2+(d-1){\rm St}} +1 \Big] \overline{U}^2 \frac{\partial p_{\rm int}  }{\partial  \rho }  \label{eq:a}  \ , \\
    b(\rho) =&   \frac{3 \ell_0^2 }{2d(d-1)(d+2)(1+2d{\rm St})(1+(d-1){\rm St})}  \Big[  \frac{4d {\rm St}}{ 2+(d-1){\rm St}} \frac{\partial }{\partial \rho } \Big( \overline{U}^2 \frac{\partial p_{\rm int}  }{\partial \rho} \Big) \nonumber \\ &+ \overline{U} \frac{\partial}{\partial \rho} \Big( \overline{U} \frac{\partial p_{\rm int}  }{\partial \rho}\Big)  \Big]   \label{eq:b}  \ ,\\
    E(\rho) = & 
    \overline{U}^{8d {\rm St}/(2+(5d-1){\rm St})}  \frac{\partial p_{\rm int} }{\partial \rho} \ . \label{eq:E}
\end{align} 
\end{subequations}
One can verify that Eqs.~\eqref{eq:P}$-$\eqref{eq:E} reduce to the overdamped (${\rm St} = 0$) case found in Ref.~\cite{Omar23b}. Specifically, the weighting factor for the overdamped case is found to be:
\begin{equation}
\label{eq:E2}
    E(\rho) = \partial p_{\rm int} /\partial \rho \ .
\end{equation}

\begin{figure}[h]
	\centering
	\includegraphics[width=.5\textwidth]{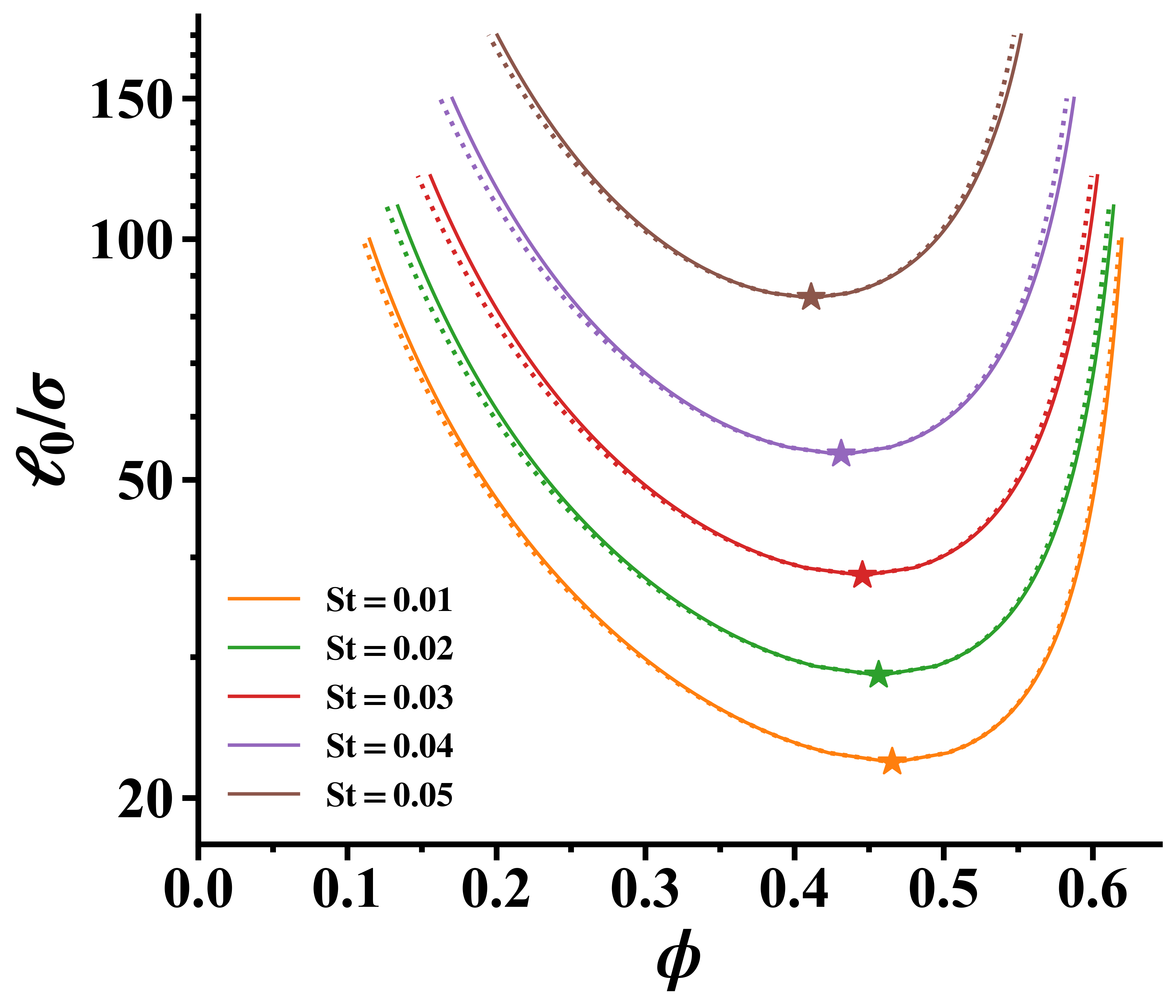}
	\caption{\protect\small{{
	Comparison of constructed binodals using our inertial mechanical theory (Eq.~\eqref{eq:E}, solid lines) and those obtained from the overdamped mechanical theory (Eq.~\eqref{eq:E2}, dotted lines).}}}
	\label{Fig:S9}
\end{figure}

Figure~\ref{Fig:S9} compares the binodals obtained for our system using the inertial criteria [Eq.~\eqref{eq:E}] and the overdamped criteria [Eq.~\eqref{eq:E2}].
When applying the inertial criteria, we identify \\ $\overline{U}=\exp{  \left[ - A \phi^B/(1- \phi/\phi_{\rm RCP})^C  \right] }$ by combining Eq.~\eqref{eq:pk_pact} and Eq.~\eqref{eq:pact}.
While the overdamped criteria provide a good approximation at the low values of inertia considered here, the inertial criteria predicts slightly higher density for both the liquid and gas phase. 
The small correction introduced by the inertial criteria is due to the fact that the correction factor in Eq.~\eqref{eq:E} does not strongly vary with $\phi$ with fixed $\ell_0/\sigma$ and ${\rm St}$ in the parameter space presented in Fig.~\ref{Fig:S9}.

\begin{figure}[h]
	\centering
	\includegraphics[width=.5\textwidth]{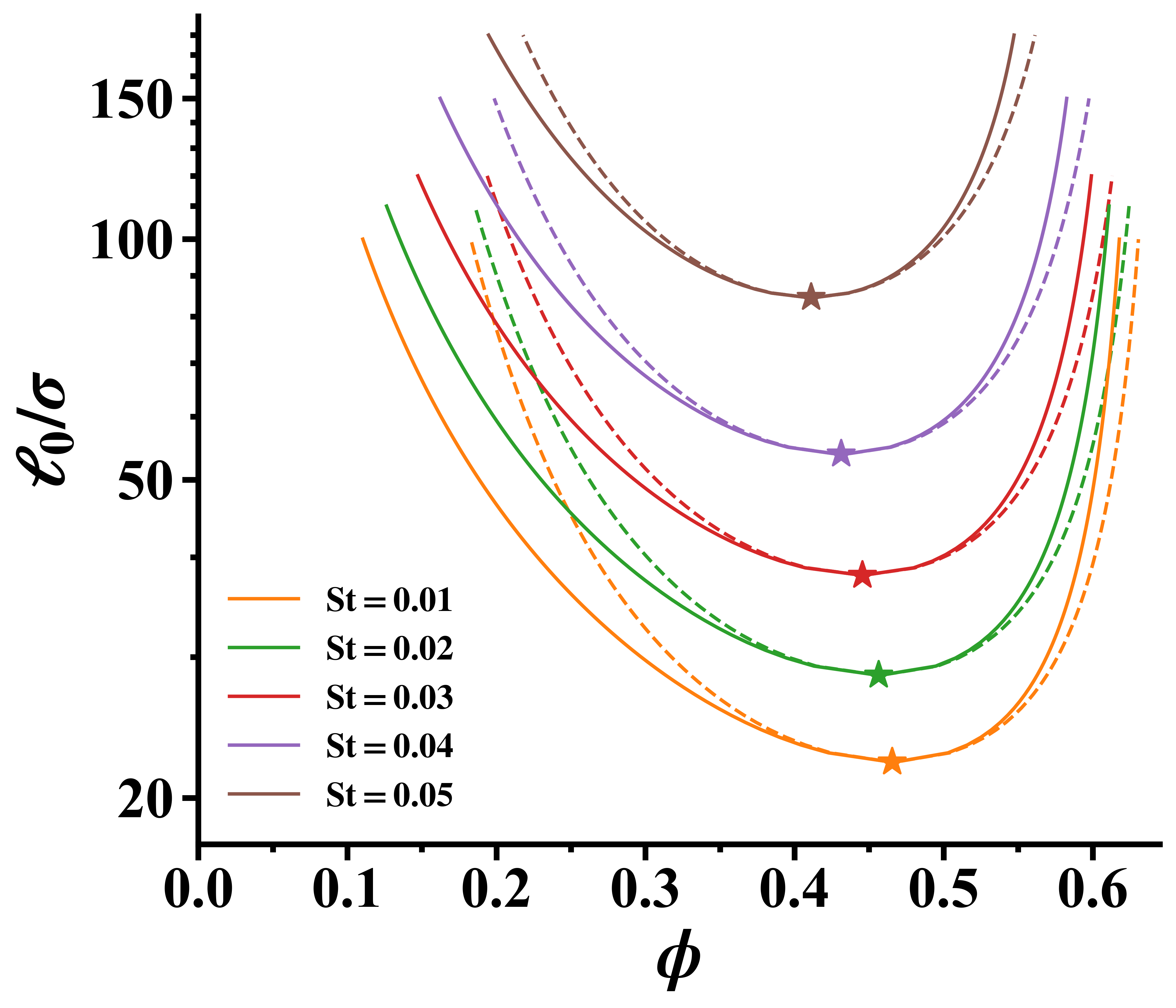}
	\caption{\protect\small{{
	Comparison of constructed binodals using our mechanical theory (Eq.~\eqref{eq:E}, solid lines) and those obtained from the equilibrium theory (an equal area construction in the $\mathcal{P}-1/\phi$ plane) (dashed lines).}}}
	\label{Fig:S10}
\end{figure}

Figure~\ref{Fig:S10} compares our mechanical theory and the equilibrium equal-area Maxwell construction in the $\mathcal{P}-1/\phi$ plane.
Consistent with Ref.~\cite{Omar23b}, we find that our mechanical theory predicts a notably smaller gas density and a slightly smaller liquid density than those predicted by the equilibrium criteria.
The resulting coexisting densities are in fact more disparate than what the equilibrium theory predicts with this trend becoming increasingly pronounced with increasing $\ell_0/\sigma$ with fixed ${\rm St}$.
We also note that the overall difference between our mechanical theory and the equilibrium equal-area construction becomes smaller with increasing ${\rm St}$, reflecting the fact that in the limit of large translational inertia, an ``effective equilibrium'' distribution is restored~\cite{Omar23a}.
\newpage

Using our theory, we can locate the critical activity $(\ell_0/\sigma)_{\rm c}$ for any given ${\rm St}$, denoted by the solid line in Fig.~3 in the main text. 
The fitted expression for the critical activity has the following form: 
\begin{equation*}
    (\ell_0/\sigma)_{\rm c} = \frac{0.04126}{(0.08624-{\rm St})^{2.2627}} + 9.2552  \,
\end{equation*}
where we can now identify that above ${\rm St}=0.08624$ MIPS ceases to occur. 
While the precise value of this critical ${\rm St}$ depends on the specific forms of equations of state, our theory qualitatively captures the fundamental observation that there exists a critical ${\rm St}$ beyond which MIPS is totally eliminated.

\section{Stability Analysis}
\subsection{Linear Stability Analysis}
Reference~\cite{Omar23b} used mechanical arguments to show that the spinodal condition for overdamped ABPs is $(\partial \mathcal{P}/\partial \rho) < 0$, where $\mathcal{P}$ is the dynamic pressure of the system. 
Here we generalize this argument to systems with inertia.
From Eq.~\eqref{eq:rho}, the evolution equation of density profile reads 
\begin{equation}
\frac{\partial \rho}{\partial t} = - \bm{\nabla} \cdot \mathbf{j}^{\rho}  \ .
\label{eq:rho2}
\end{equation}
We now need the evolution equation for number density flux $\mathbf{j}^{\rho} = \rho \mathbf{u}$. From Eq.~\eqref{eq:jrho}, this reads
\begin{equation}
    m \frac{\partial \mathbf{j}^{\rho}}{\partial t} = \bm{\nabla} \cdot \bm{\Sigma} + \mathbf{b} - \zeta \mathbf{j}^{\rho} \ .
    \label{eq:dynamics}
\end{equation}
While under steady states a system with a planar interface will satisfy $\mathbf{j}^{\rho}=\mathbf{0}$, unsteady states will permit the existence of density fluxes.
Hence we retain the drag force density in Eq.~\eqref{eq:dynamics}.
We continue to use the dynamic stress definition invoked in Eq.~\eqref{eq:dynstresscontributions}.
Note that under transient conditions, there will be additional internal body forces that cannot be readily adsorbed into the dynamic stress.
These transient internal body forces $\mathbf{b}$ may also be generated in driven systems and relax on a characteristic timescale $\tau_{\rm R}$.
We consider timescales much larger than $\tau_{\rm R}$ (and thus also the momentum relaxation time $\tau_{\rm M}$), allowing us to ignore the body force in Eq.~\eqref{eq:dynamics}.
Moreover, as we will focus on long wavelength perturbations, we omit the spatial gradient terms in the effective stress $\bm{\Sigma}$, resulting in $\bm{\Sigma} = - \mathcal{P} \mathbf{I}$.

We consider a system initially at rest $\mathbf{u} ( \mathbf{x} , t_0) = 0$ with a uniform density $\rho( \mathbf{x} , t_0) = \rho_0$. 
We now consider small amplitude perturbations to the density and velocity fields such that $\rho = \rho_0 + \delta \rho$ and $\mathbf{u}= \delta \mathbf{u}$.
Substitution of these perturbed fields into Eqs.~\eqref{eq:rho2} and~\eqref{eq:dynamics}, we have
\begin{align}
\label{eq:rho3}
\frac{\partial \delta \rho}{\partial t} + \rho_0 \bm{\nabla} \cdot \delta \mathbf{u} &= 0 \ , \\
\label{eq:dynamics2}
m \rho_0 \frac{\partial \delta \mathbf{u}}{\partial t} - \bm{\nabla} \cdot \bm{\Sigma}  + \zeta \rho_0 \delta \mathbf{u} &= 0 \ .
\end{align}
Taking a time derivative of Eq.~\eqref{eq:rho3} and substituting Eq.~\eqref{eq:dynamics2} into the resulting equation allow us to have:
\begin{equation}
\label{eq:tele}
    \frac{\partial^2 \delta \rho}{\partial t^2} + \frac{\zeta}{m}\frac{\partial \delta \rho}{\partial t}= \frac{1}{m} \left( \frac{\partial \mathcal{P}}{\partial \rho} \right)_{\rho=\rho_0} \nabla^2 \delta \rho \ ,
\end{equation}
where we have used $\bm{\nabla} \cdot \bm{\Sigma} = - (\partial \mathcal{P}/\partial \rho)_{\rho=\rho_0} \bm{\nabla} \delta \rho$.
We define the collective diffusion constant $D=(\partial \mathcal{P}/\partial \rho)_{\rho=\rho_0} / \zeta$. 
A spatial Fourier transformation of Eq.~\eqref{eq:tele} leads to:
\begin{equation}
    \frac{\partial^2 \delta \rho_k}{\partial t^2} + \frac{1}{\tau_{\rm M}} \frac{\partial \delta \rho_k}{\partial t}+\frac{Dk^2}{\tau_{\rm M}}   \delta \rho_k = 0 \ ,
\label{eq:dyna_fin}
\end{equation}
whose general solution reads
\begin{equation*}
    \delta \rho_k = A \exp \left[ - \frac{1}{2\tau_{\rm M}} \left( 1 + \sqrt{1 - 4Dk^2 \tau_{\rm M}} \right) t \right] + B \exp \left[ - \frac{1}{2\tau_{\rm M}} \left( 1 - \sqrt{1 - 4Dk^2 \tau_{\rm M}} \right) t \right] \ ,
\end{equation*}
where $A$ and $B$ are constants set by initial condition.
In the limit of large wavelengths, we have $\tau_{\rm M} \ll 1/Dk^2$ and the solution to Eq.~\eqref{eq:dyna_fin} is
\begin{equation}
    \delta \rho_k =A \exp [-Dk^2 t] + B \exp \left[ \left( Dk^2-\frac{1}{\tau_{\rm M}} \right) t\right] \ .
    \label{eq:solu}
\end{equation}
From the first term of the solution in Eq.~\eqref{eq:solu} we can read out that the density perturbations will be linearly unstable when $D<0$, recovering the spinodal condition $(\partial \mathcal{P}/\partial \rho ) < 0$. 
The second term further tells us that perturbations will also be unstable when $1/\tau_{\rm M} < Dk^2$, a condition which cannot be satisfied in the limits in which our solution is valid.

\subsection{Relative Stability of Phase-separated and Homogeneous States}
In this section, we discuss the relative stability of the phase-separated state compared to the homogeneous state within the binodal region.
The dynamical linear stability analysis made clear that, homogeneous states are unstable within the spinodal region of the phase diagram.
Therefore, we conclude it is clear phase separation is the more stable configuration (by default) within the spinodal region in comparison to the homogeneous state.

The situation is more subtle for states prepared between the binodal and spinodal where both states are stable with respect to linear perturbations. 
However, we can rely here on the wealth of simulation data in which it is observed that homogeneous states can transition to phase-separated states (often via nucleation) on measurable timescales but the reverse process has never been observed in simulation. 
From this marked discrepancy in the transition rate between two states, it is likely that in regions between the spinodal and binodal, phase separation is indeed the more probable state in comparison to a homogeneous state~\cite{Omar21}.
This intuition tacitly assumes that the forward and reverse transition pathways between these two states are identical, which is not guaranteed for active systems.
Rigorously establishing the relative state probabilities would require the use of the minimum action framework for nonequilibrium phase transitions~\cite{Zakine23} but is beyond the scope of the present work.